\documentclass[twocolumn,amsmath,amssymb,aps,superscriptaddress,longbibliography,10pt]{revtex4-2}

\usepackage{graphicx}
\usepackage{dcolumn}
\usepackage{bm}
\usepackage{color}

\usepackage{url}
\usepackage{braket}
\usepackage{comment}
\usepackage{here}
\usepackage{algorithm}
\usepackage{algpseudocode}
\usepackage{booktabs}
\usepackage[colorlinks=true,citecolor=blue,linkcolor=blue,urlcolor=blue]{hyperref}
\usepackage{lipsum}
\usepackage{chemformula}
\usepackage{tabularx,booktabs}
\usepackage[utf8]{inputenc}

\begin{document}
\title{Atomic scale structure and dynamical properties of \ch{(TeO2)_{1-x}-(Na2O)_{x}} glasses through first-principles modeling  and XRD measurements}

\author{Firas Shuaib}
\affiliation{Institut de Recherche sur les C\'eramiques (IRCER), UMR CNRS 7315, F-87068 Universit\'e de Limoges, Centre Européen de la Céramique, 12 rue Atlantis , Limoges, France.}
\author{Assil Bouzid}
\email{assil.bouzid@cnrs.fr}
\affiliation{Institut de Recherche sur les C\'eramiques (IRCER), UMR CNRS 7315, F-87068 Universit\'e de Limoges, Centre Européen de la Céramique, 12 rue Atlantis , Limoges, France.}
\author{Remi Piotrowski}
\affiliation{Institut de Recherche sur les C\'eramiques (IRCER), UMR CNRS 7315, F-87068 Universit\'e de Limoges, Centre Européen de la Céramique, 12 rue Atlantis , Limoges, France.}
\author{Gaelle Delaizir}
\affiliation{Institut de Recherche sur les C\'eramiques (IRCER), UMR CNRS 7315, F-87068 Universit\'e de Limoges, Centre Européen de la Céramique, 12 rue Atlantis , Limoges, France.}
\author{Pierre-Marie Geffroy}
\affiliation{Institut de Recherche sur les C\'eramiques (IRCER), UMR CNRS 7315, F-87068 Universit\'e de Limoges, Centre Européen de la Céramique, 12 rue Atlantis , Limoges, France.}
\author{David Hamani}
\affiliation{Institut de Recherche sur les C\'eramiques (IRCER), UMR CNRS 7315, F-87068 Universit\'e de Limoges, Centre Européen de la Céramique, 12 rue Atlantis , Limoges, France.}
\author{Raghvender Raghvender}
\affiliation{Institut de Recherche sur les C\'eramiques (IRCER), UMR CNRS 7315, F-87068 Universit\'e de Limoges, Centre Européen de la Céramique, 12 rue Atlantis , Limoges, France.}
\author{Steve Dave Wansi Wendji}
\affiliation{Université de Strasbourg, CNRS, Institut de Physique et Chimie des Matériaux de Strasbourg, UMR 7504, F-67034 Strasbourg, France.}
\affiliation{ADYNMAT CNRS consortium, F-67034, Strasbourg, France.}
\author{Carlo Massobrio}
\affiliation{Universit\'e de Strasbourg, CNRS, Laboratoire ICube UMR 7357, F-67037 Strasbourg, France.}
\affiliation{ADYNMAT CNRS consortium, F-67034, Strasbourg, France.}
\author{Mauro Boero}
\affiliation{Universit\'e de Strasbourg, CNRS, Laboratoire ICube UMR 7357, F-67037 Strasbourg, France.}
\affiliation{ADYNMAT CNRS consortium, F-67034, Strasbourg, France.}
\author{Guido Ori}
\affiliation{Université de Strasbourg, CNRS, Institut de Physique et Chimie des Matériaux de Strasbourg, UMR 7504, F-67034 Strasbourg, France.}
\affiliation{ADYNMAT CNRS consortium, F-67034, Strasbourg, France.}
\author{Philippe Thomas}
\affiliation{Institut de Recherche sur les C\'eramiques (IRCER), UMR CNRS 7315, F-87068 Universit\'e de Limoges, Centre Européen de la Céramique, 12 rue Atlantis , Limoges, France.}
\author{Olivier Masson}
\affiliation{Institut de Recherche sur les C\'eramiques (IRCER), UMR CNRS 7315, F-87068 Universit\'e de Limoges, Centre Européen de la Céramique, 12 rue Atlantis , Limoges, France.}

\date{\today}

\begin{abstract}
\noindent\normalsize{We resort to first-principles molecular dynamics, in synergy with
experiments, to study structural evolution and Na$^+$
cation diffusion inside \ch{(TeO2)$_{1-x}$-(Na2O)$_{x}$} (x = 0.10-0.40) glasses.
Experimental and modeling results show a fair quantitative agreement in terms of total X-ray structure factors and pair distribution functions, 
thereby setting the ground for a comprehensive analysis of the glassy matrix evolution. 
We find that the structure of \ch{(TeO2)$_{1-x}$-(Na2O)$_{x}$} glasses deviates drastically from that of pure \ch{TeO2} glass. 
Specifically, increasing the \ch{Na2O} concentration leads to a reduction of the coordination number of Te atoms, reflecting the occurrence 
of a structural depolymerization upon introduction of the \ch{Na2O} modifier oxide. The depolymerization phenomenon is ascribed to the transformation 
of Te-O-Te bridges into terminal Te-O non bridging oxygen atoms (NBO). 
Consequently, the concentration of NBO increases in these systems as the concentration of the modifier increases, accompanied by
a concomitant reduction in the coordination number of Na atoms. 
The structure factors results show a prominent peak at $\sim1.4$ \AA\, that becomes more and more pronounced as the \ch{Na2O} concentration increases. 
The occurrence of this first sharp diffraction peak is attributed to the growth of Na-rich channels inside the amorphous network,
acting as preferential routes for alkali-ion conduction inside the relatively stable Te-O matrix. These channels enhance the ion mobility.}
\end{abstract}

\maketitle

\section{Introduction}
Tellurite glasses have been the target of considerable research 
efforts \cite{torzuoli2020enhanced,kavakliouglu2015teo2,bouzid2022buckingham, stanworth1952tellurite,barney2013terminal, el1989theoretical, el1998tellurite, sarjeant1967new, rajendran2003characterisation, moraes2010relation} thanks to their advantageous properties, such as the high refractive index (1.8–2.3), high dielectric constant (13–35), and outstanding non-linear optical properties \cite{kim1993linear, nasu1990third, lim2004structure}.
Moreover, these glasses are highly regarded as excellent materials for accommodating lasing ions due to their ability to provide a low phonon energy environment (600–850 $\hbar$w/cm), which effectively minimize non-radiative losses \cite{durga2002optical, durga2002physical}.\\

Pure tellurium oxide is a conditional glass former and can be produced only in small amounts 
through fast quenching from the melt state\cite{barney2013terminal} as this material lacks stability and is prone to rapid crystallization. This is mostly due to the existence of stereoactive Te lone electron pair (when bonded to O) that limits the  structural rearrangements required for achieving a glassy state \cite{tagg1995structure}.
The production of \ch{TeO2} glasses under normal quenching conditions, thus, requires the addition of a modifier oxide. For example, stable binary glasses can be obtained by mixing \ch{TeO2} with a modifier oxide  such as \ch{Tl2O} \cite{bouzid2022buckingham}, Na$_{2}$O \cite{kavakliouglu2015teo2}, \ch{ZnO} \cite{de2021new}, or \ch{Li2O} \cite{rammah2020investigation}. This procedure achieves stable glassy systems and enables the control of their physical and chemical properties \cite{naresh2020influence} by monitoring the concentration of the added modifier oxide. Hence, finding suitable glassy compositions for a particular application requires a fine understanding of the influence of the modifier oxide on the structural properties of the host material \cite{el1998tellurite,stepien2011development}.\\

Among the various modifier oxides, \ch{Na2O} stands as an interesting candidate to investigate the correlation between the chemical composition of the glass and its structural and dynamical properties. Actually, \ch{Na2O} allows to decrease the glass transition temperature and viscosity of tellurite glasses, at the same time it stabilizes the \ch{TeO2} network and triggers ionic conductivity suitable for batteries applications \cite{ccelikbilek2013glass,desirena2009effect,tincher2010viscosity}.
Despite these advantages, the precise role of the Na atom in the network connectivity is not well understood and call for a deeper investigation of structure-composition-properties relationships.

In literature, various approaches have been used to investigate the structure on binary alkali tellurite glasses, including neutron diffraction \cite{barney2013terminal, neov1994neutron}, X-ray diffraction combined with reverse
Monte Carlo (RMC) modelling \cite{zwanziger1997sodium, mclaughlin2001structure, hoppe2005structure}, Raman spectroscopy \cite{garaga2017short}, and magic angle spinning (MAS)-nuclear magnetic resonance (NMR) \cite{sakida1999part, sakida1999part2}. Nevertheless, the diverse nature of the tellurium-oxygen structural units poses challenges in accurately describing the composition dependent alteration in the structural glass local environment and overall connectivity. According to Neov \textit{et al.} \cite{neov1994neutron}, neutron diffraction  indicates that in \ch{TeO2-M2O} (M = Li, Na, K or Rb) glasses with high \ch{TeO2} concentration ($<$ 4 mol\% modifier), the tellurium main structural unit is a four-fold coordinated disphenoid \ch{TeO4}, similar to that observed in paratellurite ($\alpha$-\ch{TeO2}) polymorph. In addition, they revealed that by decreasing the \ch{TeO2} concentration (increasing the MO concentration), the Te local environment undergoes a modification to a three-fold coordinated structure through the elongation of one \ch{Te-O} linkage. In many studies on alkali-tellurite glasses, in particular those based on RMC modeling \cite{sakida1999part,mclaughlin2000structure,mclaughlin2001structure,heo1992spectroscopic}, it has been proposed that the tellurium polyhedra present in the glasses are the same as the five polyhedra found in the well-known crystalline phases, in variable proportions \cite{mclaughlin2000structure,mclaughlin2001structure,sakida1999part}. McLaughlin \textit{et al.} \cite{mclaughlin2000structure} showed that the abundance of each individual tellurite polyhedra is charge-dependent 
on the $Q^{n}_{m}$ units, where $m$ is the coordination number and $n$ denotes the number of bridging oxygen atoms. Specifically, as the modifier oxide concentration increases, the quantity of uncharged tellurite polyhedra ($Q^{4}_{4}$, $Q^{2}_{3}$) decreases, while the quantity of charged polyhedra ($Q^{3}_{4}$, $Q^{1}_{3}$, $Q^{0}_{3}$) increases \cite{mclaughlin2001structure}.\\

On the experimental side, extensive investigations have been conducted on the synthesis, identification of new crystalline compounds, glass formation domains, and structural analysis of binary \ch{Na2O-TeO2} glasses using various approaches 
\cite{sekiya1989normal,sekiya1992raman,sakida1999part,mclaughlin2000structure,mclaughlin2001structure,zwanziger1997sodium,tagg1995structure}. 
The phase diagram of this binary system in the 10 $\leq$ x $\leq$ 50 mol\% \ch{Na2O} compositional range was reported for the first time in the literature by Troitskii et al \cite{troitskii1967izv, kavakliouglu2015teo2}. 
Mochida \textit{et al.} reported the first findings about the glass forming abilities of the (100-x)\ch{(TeO2)-x(Na2O)} system. In their work, the glass formation domain was established as 10 $\leq$ x $\leq$ 46.5 mol\% \cite{mochida1978properties, barney2015alkali, kavakliouglu2015teo2}. 
More recently, Kutlu \textit{et al.} \cite{kavakliouglu2015teo2} reported that glass formation range is established as 7.5 $\le$ x $\le$ 40 mol\%, and revealed the network-modifying effect of \ch{Na2O}, leading to a decrease in the glass transition temperature and the formation of a less dense glass structure with increasing \ch{Na2O} content. 

Focusing on the structure, Sekiya \textit{et al.}\cite{sekiya1992raman} found that \ch{TeO2} glasses with a small quantity of \ch{Na2O} (up to 20 mol\%) form a random network of \ch{TeO4} disphenoids ($Q_4^4$) and highly distorted disphenoid with three short Te-O bonds and one longer bond leading to \ch{TeO_{3+1}} polyhedra, with one non-bridging oxygen (NBO). As the \ch{Na2O} content was increased, the fraction of NBO increases and \ch{TeO3} trigonal pyramids become the main building bloc of the glassy network. 
X-ray diffraction experiments complemented with $^{\rm 23}$Na NMR spectroscopy\cite{tagg1995structure}, showed that the coordination number of Na atoms drops from around 6 for $x \le$ 0.20 to about 5 when $x = 0.35$ and it was concluded that the 5-fold coordinated environment is more representative of the glass\cite{tagg1995structure}. 
Furthermore, it was observed that the sodium environment at $x = 0.20$ is significantly distinct between the crystal and glass forms.  
Zwanziger \textit{et al}\cite{zwanziger1997sodium} showed that at low Na concentrations, the distribution of sodium ions in the glassy matrix seems to be random. However, at larger concentrations, there is a notable intermediate range order (IRO), which is particularly evident at x = 0.20. 

Accurately accounting for the electronic structure of the glasses should clarify the  properties of the \ch{Te-O} bond and how they are influenced by the presence of modifier ions leading to a precise picture on the network connectivity and Na ionic dynamics. The investigation of ionic migration mechanisms requires the exact description of the atomic structure. At this level, molecular dynamics (MD) \cite{allen2017computer} can (i) provide a deeper understanding of the atomistic structure, thereby complementing experimental efforts, and can (ii) offer quantitative details on the ionic transport mechanisms, when using an accurate interaction model. Specifically, first-principles molecular dynamics FPMD approach \cite{massobrio2015molecular,massobrio2022structure,levchenkotheory}, explicitly taking into account the electronic structure of the system, enables the accurate modeling of glassy structures. Nevertheless, the process of deducing the diffusion mechanisms of alkali ions in such glassy systems remains to some extent complex.

In this work, supported by experimental investigations, we use first-principles molecular dynamics (FPMD) methods \cite{massobrio2015molecular,massobrio2022structure,levchenkotheory} to generate model systems of several glass compositions where the common template is the \ch{(TeO2)$_{1-x}$-(Na2O)$_{x}$} system. 
Special attention is given to the changes occurring in both the structure of the network and the related dynamical properties when the \ch{Na2O} content is schanged. Moreover, we investigate the infulence of the local environment of the alkali cations (i.e. the link between the \ch{Na–O} coordination polyhedra) in the diffusion mechanism of Na$^+$ its role in the ionic conduction in this type of materials. 

The paper is organized as follows: Computational details are presented in section 1 where we provide a description of the FPMD methodology, \ch{(TeO2)$_{1-x}$-(Na2O)$_{x}$} model generation process and a description of the X-ray structure factor and pair distribution function calculation. This section includes also a dedicated part that provides the definitions of the Wannier centers based analysis, calculated structural and dynamical quantities. Section 2 presents the synthesis and characterization protocols. The results are outlined in section 3 and divided into two subsections. The first subsection  focuses on structural characterization of our samples and the second one focuses on the dynamical features. The conclusions of our work are presented in section 4.\\
 
 \section{Computational methods}
 \subsection{First-principles molecular dynamics simulations}
The electronic structure calculations were carried out within the density functional theory (DFT) framework as implemented in the CP2K code \cite{hutter2014cp2k}. 
A mixed Gaussian Plane Waves (GPW) basis set was used to represent the electronic structure of the Kohn-Sham functional. 
Specifically, valence orbitals
were described by atom-centered Gaussian-type triple-$\zeta$ fucntions \cite{peintinger2013consistent}, while the electron density was 
expanded on an auxiliary plane-wave basis set with an energy cutoff of 600 Ry (8163.4 eV).
The Brillouin zone sampling was restricted to the $\Gamma$ point. 
Analytic Goedecker-Teter-Hutter norm-conserving pseudopotentials were used to describe core-valence electronic interactions \cite{goedecker1996separable}.
The exchange-correlation functional used in our simulations was the one proposed by Perdew, Burke, and Ernzerhof (PBE) \cite{perdew1998perdew}. 

The dynamical simulation protocol was the Born-Oppenheimer molecular dynamics (BOMD) \cite{waller1956dynamical,combes1981born} as implemented in CP2K code \cite{hutter2014cp2k}. The equations of motion were numerically integrated with a time step of $\Delta$t = 1 fs, ensuring optimal conservation of the total energy at least on the ps time scale of the simulations. FPMD simulations were carried out in the NVT ensemble (constant number of particles, volume and temperature) and the temperature control was operated by a Nosé-Hoover thermostat chain \cite{nose1984unified,martyna1992nose}.\\

\subsection{Model generation protocol}
 
Four \ch{(TeO2)_{1-x}-(Na_{2}O)_{x}} glasses with x= 0.10, 0.1875, 0.30, and 0.40 were produced through quenching from the melt. The initial configuration of each system is made of 480 atoms randomly placed in a cubic simulation cell, of side length equal to 19.86, 19.94, 20.01 and 20.02 {\AA}, respectively, and corresponding to the measured experimental densities provided in table [\ref{table0}]. Periodic boundary conditions were applied for all models. Glassy configurations are obtained according to the following thermal cycle: T = 2000 K (10 ps), T = 1000 K (25 ps), T = 750 K (25 ps), T = 500 K (25 ps), and finally T = 300 K (25 ps). At the highest temperature of the thermal cycle  (T = 2000 K) all the atomic species (\ch{Te}, \ch{O}, and \ch{Na}) were found to reach high mobility, characterized by a liquid-like diffusion  coefficients ($\sim$10$^{-5}$cm$^{2}$s$^{-1}$) ensuring a complete loss of the initial state memory. In this work the last 20 ps of the trajectory at T = 300 K were used to compute the statistical averages of the structural properties for all the \ch{(TeO2)_{1-x}-(Na_{2}O)_{x}} glasses.\\

\subsection{X-ray structure factor and pair distribution function calculation}
Insights about the structural evolution of \ch{(TeO2)_{1-x}-(Na_{2}O)_{x}} glasses with varying concentration x can be obtained by investigating both the reciprocal space properties through the X-ray structure factor S$_{X}$(q) and the real space pair distribution function (PDF).
The structure factor is defined by:

\begin{eqnarray}\label{eq2}
S_{X}(q) = 1 + \sum_{\alpha=1}^{3}\sum_{\beta = 1}^{3}\frac{c_{\alpha}c_{\beta}f_{\alpha}(q)f_{\beta}(q)}{<f(q)>^{2}}\Big[S_{\alpha\beta}^{FZ}(q)-1\Big], \label{eq1}
\end{eqnarray}

where, the chemical species (\ch{Te}, \ch{O}, or \ch{Na}) are denoted $\alpha$ and $\beta$, $q$ is the magnitude of the scattering vector, $f_{\alpha}(q)$ and $c_{\alpha}$ are the scattering factor and the atomic concentrations of species, respectively and $<f(q)>$ = $\sum_{\alpha}{c_{\alpha}f_{\alpha}(q)}$ is the average scattering factor.
$S_{\alpha\beta}^{FZ}(q)$ defines the Faber–Ziman (FZ) partial structure factors \cite{faber1965theory}. A direct access to the $S_{\alpha\beta}^{FZ}$ partial structure factors on the equilibrium trajectory can be obtained via Fast Fourier transform (FFT) of the real space atomic pair distribution functions $g_{\alpha\beta} (r)$ as follows:

\begin{eqnarray}
S_{\alpha\beta}^{FZ}(q) = 1 + \frac{4\pi\rho_{0}}{q}\int_{0}^{\infty}r\Big[ g_{\alpha\beta} (r)-1\Big] \sin(qr)dr.
\end{eqnarray}

The X-ray PDF ($G_{X}(r)$) is usually defined as follows:

\begin{eqnarray}
G_{X}(r) \sim \sum_{\alpha=1}^{3}\sum_{\beta = 1}^{3}\frac{c_{\alpha}c_{\beta}f_{\alpha}(q_{0})f_{\beta}(q_{0})}{|<f(q_{0})>^{2}|}\Big[g_{\alpha\beta}(r)-1\Big].
\end{eqnarray}

where, $g_{\alpha\beta}$ is the partial pair distribution function for species $\alpha$ and $\beta$ and where the scattering factors are evaluated at an arbitrary $q_0$ typically set to $q_0 = 0$. To ensure a proper comparison to experiments, we instead employed the exact expression for ${\rm G}(r)$ from Ref.[\citenum{masson2013exact}], which involves a weighted linear combination of modified partial pair correlation functions. The determination of the weights is based on the mean values of the Faber-Ziman factors over the range of the considered reciprocal space. Subsequently, the PDF is transformed to reciprocal space and back transformed to real space using the experimental Q$_{max}$ value. This procedure, ensures a fair comparison of the calculated PDF to the experimental one by guaranteeing similar data treatment.

\subsection{Wannier centers based analysis}\label{subs2}

The network connectivity and the structural units of the amorphous systems were investigated using the maximally localized Wannier functions $w_{n}$(r) formalism (MLWFs) \cite{resta1999electron, marzari2012maximally}.
Within this formalism, a Wannier center (W) gives access to the most probable spatial localization of the shared or lone pair electrons. In practice, Wannier orbitals ($w_{n}$(r)) and their corresponding centers are obtained by an on the fly unitary transformation of the Kohn–Sham orbitals $\psi$$_{i}$(r) under the constraint of minimizing the spatial extension (spread, $\Omega$) of the resulting $w_{n}$(r) as follows:

\begin{eqnarray}
 \Omega = \sum_{n} \Big(\big< w_{n}|r^{2}|w_{n}\big>\Big)- \Big(\big< w_{n}|r|w_{n}\big>^{2}\Big) \label{eq0}
\end{eqnarray}

The resulting center represents the localization of two electrons and indicates their average position, makes it possible to define chemical bonds and lone pair electrons. Specifically, two different Wannier centers were found for each chemical species (\ch{Te}, \ch{O}, or \ch{Na}). \ch{W^{B}} and \ch{W_{$\alpha$}^{LP}} stand for centers corresponding to chemical bonds and lone-pair electrons associated to atom $\alpha$, respectively. 

Within this formalism, if two atoms $\alpha$ and $\beta$ situated at a distance $d_{\alpha\beta}$ and sharing a Wannier center located at distances $d_{\alpha W}$ and $d_{\beta W}$ satisfy the inequality $|d_{\alpha\beta}-d_{\alpha W}-d_{\beta W}| \le 0.05$ {\AA}, they are considered to be bonded. A tolerance of 0.05 {\AA} is taken into consideration to account for the deviations in the spatial localization of the center. Additionally, it has been shown that it is necessary to define a cutoff angle between \ch{W_{Te}^{LP}},\ch{Te}, and \ch{O} to be $\ge$ 73$^\circ$ in order to dismiss lone pair Wannier centers that may arise at distances that fulfill the bonds inequality requirement but do not correspond to bonds \cite{raghvender2022structure,gulenko2014atomistic}. 

Following this procedure, O atoms are labeled as bridging (BO, e.g. forming \ch{Te-O-Te} bridges) and non bridging (NBO, e.g. forming \ch{Te-O} terminal bonds) based on their bonding nature with Te atoms. Consequently, one can achieve a counting of the number of Te-O bonds, thereby reducing the effects of selecting a fixed bond length cutoff and enabling a more accurate estimation of the coordination number of Te as well as a decomposition of the local environment around Te and O atoms.
In this work, Wannier functions are computed on top of 100 configurations selected along the last 20 ps of the trajectory at T = 300 K. 

\subsection{Dynamical simulations at finite temperatures}

FPMD simulations enable the evaluation of the transport characteristics of mobile ions in disordered materials. 
The atomic mean-square displacements (MSD) were collected for all chemical species over a period of 30 ps at various temperatures (T = 900 K, T = 1100 K, T = 1200 K, and T = 1400 K) starting from the equilibrated systems at T = 300 K.

The mean square displacement of a given chemical species $\alpha$ (MSD$_{\alpha}$) is calculated as follows:

\begin{equation}
    \Big < r_{\alpha}^{2} (t) \Big > = \text{MSD}_{\alpha} (t) = \Bigg < \frac{1}{N_{\alpha}}\sum_{i=1}^{N_{\alpha}} | r_{i}(t) - r_{i}(0)|^{2} \Bigg > = \Bigg < \sum_{i} (\Delta R_{i} (t))^{2} \Bigg >, 
    \label{d1}
\end{equation}

where, the sum runs over all atoms of type $\alpha$, and $\Big < \sum_{i} (\Delta R_{i} (t))^{2} \Big >$ is the average MSD over a given time $t$. The diffusivity $D_{\alpha}$ of chemical species $\alpha$ is determined by analyzing the MSD in log-log plots using the time-dependent slope function $\beta(t)$ that allows to disentangle the ballistic and diffusive regimes efficiently: 
\begin{eqnarray}
    \beta(t) = \frac{\text{d} \,\, \text{log} \big < r_{\alpha}^{2} (t) \big >}{\text{d} \,\, \text{log} \, t}.
    \label{d2}
\end{eqnarray} 

For extremely short time, $\beta(t)$ is predicted to be equal to two, indicating optimal ballistic motion of the ions  \cite{huang2011direct}. At long simulation times, $\beta(t)$ should approach a value of one, indicating a diffusive regime where the mean square displacement increases linearly as a function of time. 
We note that at low temperatures, the beta profile becomes noisy, which reflects the fluctuations of the mean square displacement and the limited diffusivity.
This analysis is essential for determining the successful demonstration of diffusive behavior and pinpointing the specific part of the MSD plot necessary for extracting diffusion coefficients \cite{habasaki2020molecular, pham2023unveiling}. In this work, the lower and upper limits of the actual diffusive regime is based on the interval between the point when $\beta(t)$ achieves a value of one (lower boundary) and the point where the trajectory length is 10\% less than the total length (upper boundary) \cite{pham2023unveiling}. 
In the diffusive regime, the ionic diffusion coefficients (tracer diffusivity) are determined using the Einstein relation \cite{frenkel2001understanding, allen2017computer} given by:

\begin{eqnarray}
    D_{\alpha}^{*} = \frac{1}{6}\lim_{t \to \infty} \frac{\text{d}\big < r_{\alpha}^{2} (t) \big >}{\text{d} t} \sim \frac{\sum_{i}\big<\Delta R_{i}^{2}\big >}{6} \label{d2}.
    \label{d3}
\end{eqnarray}

Once the diffusive regime is achieved, the Arrhenius equation can be applied to obtain the activation energy $E_{a}$ barrier for diffusion (conductivity) by fitting the $\text{log}D_{\alpha}$ (or $\text{log}\sigma_{\alpha}$) $vs.$ $1/T$ data,

\begin{eqnarray}
    D_{\alpha}^{*} = D_{0}\exp\Big (-\frac{\Delta E_{a}}{K_{B}T}\Big ).
    \label{d4}
\end{eqnarray}

The estimation of the ionic conductivity $\sigma_{\alpha}$ from the tracer diffusivity $D_{\alpha}^{*}$ can be achieved using the Nernst–Einstein relation:
\begin{eqnarray}
    \sigma_{\alpha}^{*} = D_{\alpha}^{*} \frac{(Z_{\alpha}e)^{2}}{Vk_{B}T} = \frac{(Z_{\alpha}e)^{2}}{6Vk_{B}T}\sum_{i} \big< \Delta R_{i}^{2}\big >,
    \label{d5}
\end{eqnarray}
where, $Z_{\alpha}$ is the nominal charge of species $\alpha$, e is charge unit, $k_{B}$ is the Boltzmann constant, $T$ is temperature and $V$ is the total volume of the model system.

We observe that equation \ref{d5} provides the ideal ionic conductivity ${\sigma_{\alpha}^{*}}$ that does not account for correlation effects in ion motion within the glass. This relationship typically assumes a Haven ratio close to 1, thereby ${\sigma_{\alpha}^{*}}$ is the lower bound conductivity that can be calculated. This assumption is justified by the overall low ionic conductivity in glasses that hinders the proper sampling of the correlated motion during short simulation times. In addition, despite being low, FPMD overestimates the room temperature ionic conductivity by several orders of magnitude \cite{he2018statistical,sasaki2025constant}. As such, assuming a Haven ratio of 1 is reasonable and usual choice to achieve a qualitative comparison between the trends of the measured and the calculated ionic conductivities.

\section{Glass synthesis and characterizations}

\ch{(TeO2)_{1-x}-(Na_{2}O)_{x}} glass compositions  with x=0.1, 0.2 and 0.3 were prepared from mixture of high purity raw materials,  \ch{TeO2} (Todini, 99.9\%) and \ch{Na2CO3} (Strem Chemicals, 99.95\%) and melted at 850°C in a platinum crucible for 30 minutes with two intermediate stirring. The melt was then poured in a 10 mm diameter mold and annealed T$_{\rm g}$-10°C. We note that an attempt to prepare glass at x=0.4 led to partially crystallized samples. 

DSC measurements were performed using a TA Instrument (AQ1000) with a heating rate of 10°C/min. Table \ref{table0} summarizes the characteristic temperatures i.e. the glass transition temperature, T$_{\rm g}$, the temperature of first crystallization, T$_{\rm x1}$ and the stability of the glass through the $\delta$T=T$_{\rm x1}$-T$_{\rm g}$ criterion. In addition, we report on Table \ref{table0}, densities measured on the glass powder using helium pycnometer (AccuPyc, Micromeritics) in a 1 cm$^3$ cell.

\begin{table}[h!]
 	\begin{center}
    \begin{small}
    \scalebox{0.9}{
 		\begin{tabular}{ccccc}
                \toprule
Composition (mol \%)	& T$_g$ (°C)	&T$_{x1}$ (°C) & $\delta$T (°C)&	Density  (gcm$^{-3}$) \\
\hline
x = 0.1	&285	&321	&36	&5.08 \\
x = 0.2	&253	&385	&132	&4.73 \\
x = 0.3	&226 &312	&86	&4.32 \\
\hline      
        \end{tabular}}
        \end{small}
    \caption{Measured glass transition temperature (T$_g$), crystallization temperature T$_{x1}$, $\delta$T=T$_{x1}$-T$_g$ and density on \ch{(TeO2)_{1-x}-(Na_{2}O)_{x}} with x=0.1, 0.2 and 0.3. The uncertainties are ± 1°C for temperature measurements and ± 0.01 gcm$^{-3}$ for density measurements.}
     \label{table0}
    \end{center}  
\end{table}

The experimental total X-ray structure factor S($k$) and reduced total pair distribution functions G($r$) of the \ch{(TeO2)_{1-x}-(Na_{2}O)_{x}} systems were determined through X-ray total scattering following procedures similar to those employed in Ref.[\citenum{micoulaut2023quantitative}].  
X-ray scattering measurements were conducted at room temperature using a specialized laboratory setup based on a Bruker D8 Advance diffractometer. This instrument was equipped with a silver sealed tube ($\lambda$ = 0.559422~\AA) and a rapid LynxEye XE-T detector to enable data collection with good counting statistics up to a large scattering vector length of 21.8~\AA$^{-1}$.
Approximately 20~mg of each sample's powder were placed in a thin-walled (0.01~mm) borosilicate glass capillary with a diameter of about 0.7~mm to limit absorption effects. The $\mu$R values (where R is the capillary radius and $\mu$ is the sample’s linear attenuation coefficient) were estimated based on precise measurements of the mass and dimensions of the samples.
The raw data were corrected, normalized, and Fourier transformed using custom software~\cite{PYTSREDX} to obtain the reduced atomic pair distribution functions G($r$). Corrections accounted for capillary contributions, empty environment, Compton and multiple scatterings, absorption, and polarization effects. The necessary X-ray mass attenuation coefficients, atomic scattering factors, and Compton scattering functions for data correction and normalization were calculated from tabulated data provided by the DABAX database~\cite{DabaxFiles}. Absorption corrections were evaluated using a numerical midpoint integration method, where the sample cross-section was divided into small subdomains, following a method similar to that proposed by A. K. Soper and P. A. Egelstaf~\cite{soper1980multiple}.\\

The conductivity was determined on the specific compositions \ch{(TeO2)_{0.9}-(Na_{2}O)_{0.1}} and  \ch{(TeO2)_{0.7}-(Na_{2}O)_{0.3}} by Electrochemical Impedance Spectroscopy (EIS) measurements with a Solartron 1260 Impedance/Gain-Phase Analyzer at frequencies ranging from 5 MHz to 1 Hz, and a voltage amplitude of 3 V. Impedance measurements were carried out during a heating stage from room temperature to about 20°C below the glass transition temperature (T$_{\rm g}$). The electronic conductivity (S/cm) was calculated using the equation $\sigma$ = L/(R.S) where L is the pellet thickness (cm), R is the overall resistance ($\Omega$), and S is the area of the pellet (cm$^2$). 

\section{Results and discussion}

\subsection{Structural properties}

\subsubsection{Structure factor analysis}

\begin{figure}[!h]
	\centering 
	\includegraphics[width=0.99\linewidth,keepaspectratio=true]{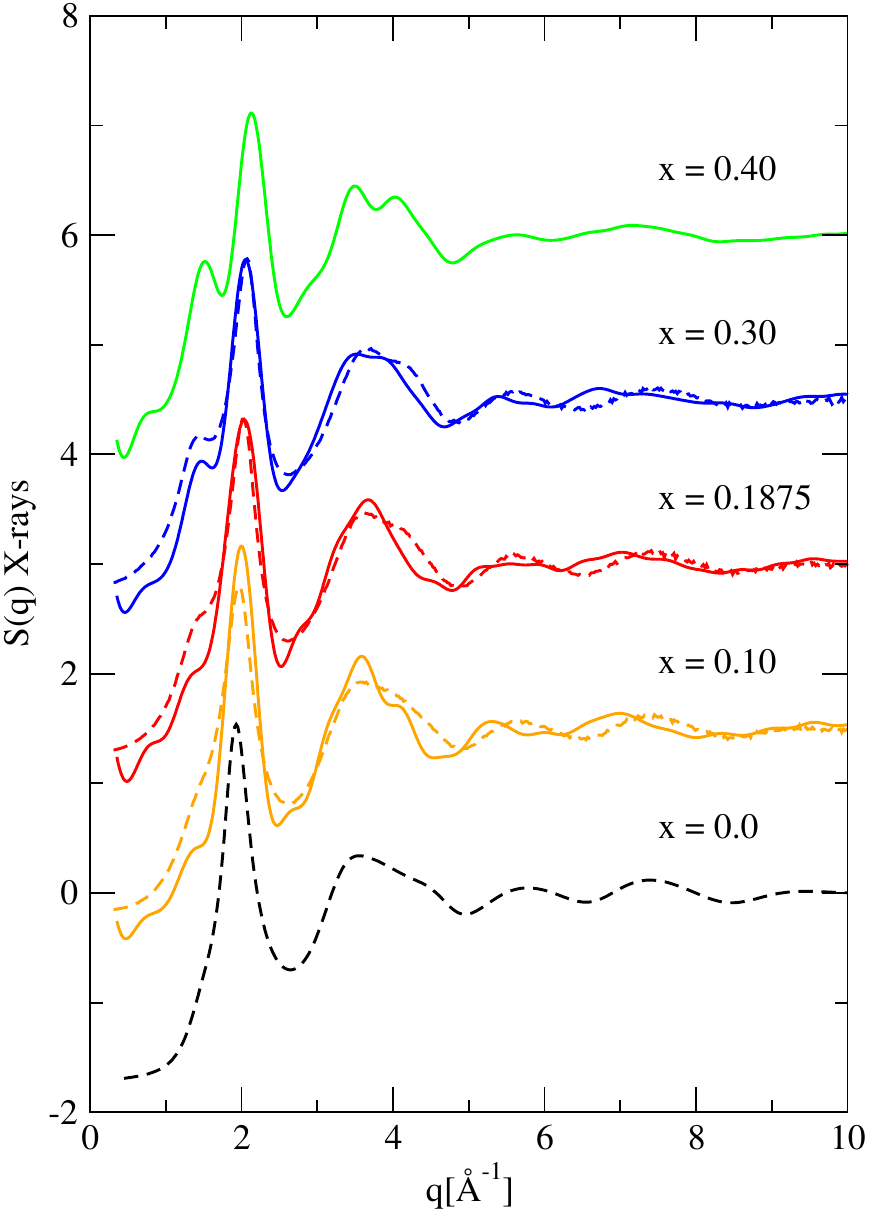}
	\caption{The experimental (dashed lines) X-ray structure factors $S_{X}(q)$ of \ch{(TeO2)_{1-x}-(Na_{2}O)_{x}} glasses with x= 0.0, 0.10, 0.1875 and 0.30 compared to the calculated total X-ray structure factors obtained via Fourier transform of the pair correlation functions in the real space (solid lines). The total X-ray structure factors of pure \ch{TeO2} taken from Ref.[\citenum{raghvender2022structure}] is also added. Vertical shifts were applied for clarity.}
	\label{fig1}
\end{figure}

The calculated X-ray structure factors S$_{X}$(q) for \ch{(TeO2)_{1-x}-(Na_{2}O)_{x}} glasses are displayed in Fig.[\ref{fig1}] and compared to their experimental counterpart for x = 0.10, 0.1875, and 0.30. In addition, data computed on a pure \ch{TeO2} glass from Ref.[\citenum{raghvender2022structure}] is also displayed.  
Overall, there is a good agreement between the calculated and the measured structure factors for all the concentrations of \ch{Na2O} and over the whole reciprocal space range. One observes a dominant peak at about 1.9 \AA$^{-1}$, whose position shifts towards larger $q$ values when increasing the \ch{Na2O} content. This dominant peak is followed by a broader peak at $\sim$ 3.6 \AA$^{-1}$ that shows less sensitivity to the \ch{Na2O} concentration. 

\begin{figure}[!h]
	\centering
	\includegraphics[width=0.99\linewidth,keepaspectratio=true]{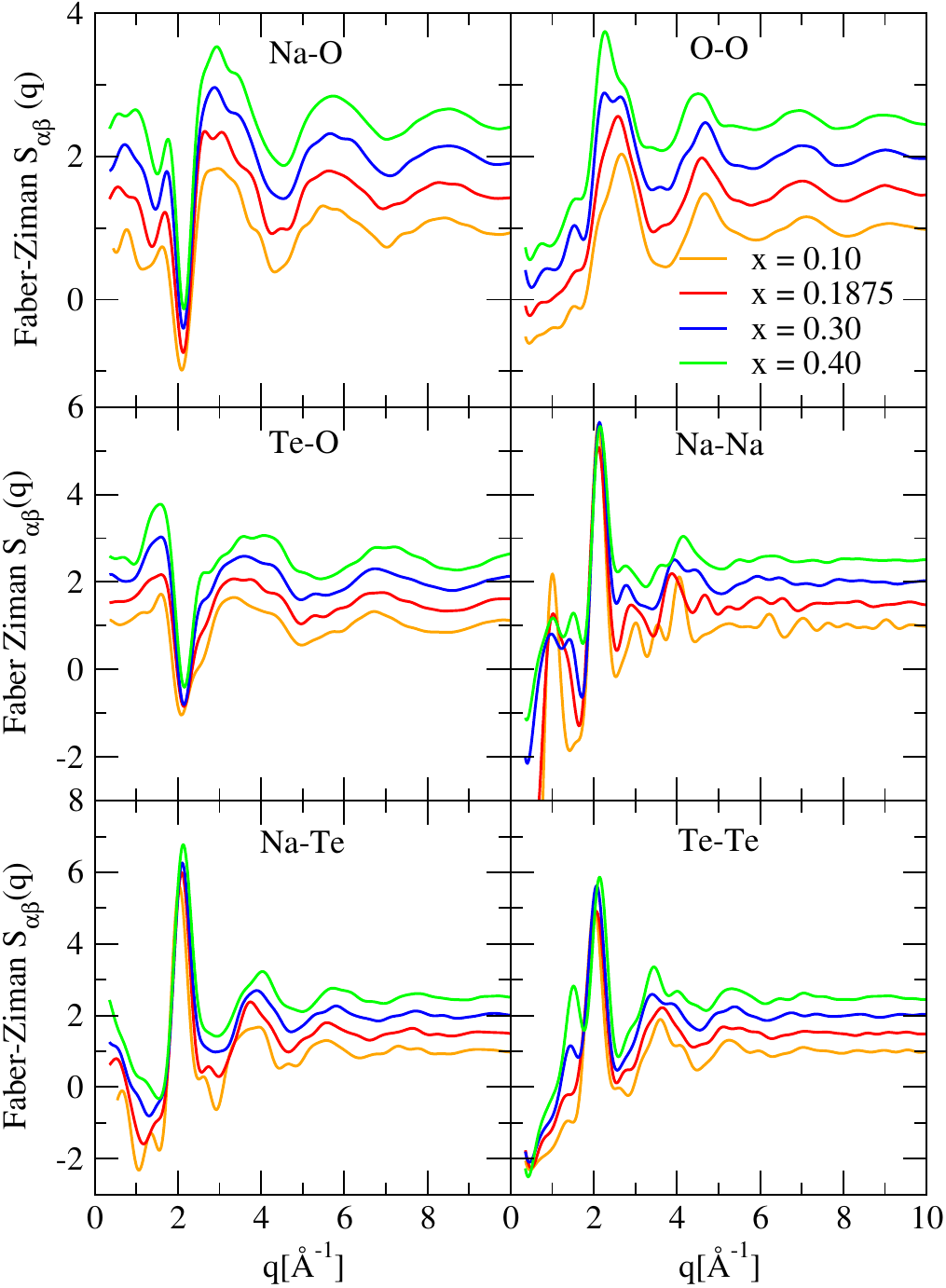}
	\caption{Comparison between the calculated Faber–Ziman partial structure factors as a function of  concentration x for the \ch{(TeO2)_{1-x}-(Na_{2}O)_{x}} glasses with x= 0.10, 0.1875, 0.30 and 0.40. Vertical shifts are applied for clarity.} 
		\label{fig2}
\end{figure}

Interestingly, one can notice a growing peak around 1.4 \AA$^{-1}$ with increasing \ch{Na2O} concentration. This peak, corresponding to the so called first sharp diffraction peak (FSDP), is related to the intermediate range order (IRO) in the glass. The Faber–Ziman (FZ) partial structure factors are shown in Fig.[\ref{fig2}]. In sodium tellurite systems, \ch{Te} features the highest scattering factor ($f_{\ch{Te}}$ (q = 0) $\simeq$ Z$_{\ch{Te}}$ = 52 compared with $f_{\ch{Na}}$ (q = 0) $\simeq$ Z$_{\ch{Na}}$ = 11 and $f_{\ch{O}}$ (q = 0) $\simeq$ Z$_{\ch{O}}$ = 8). Therefore, it is expected that $S_{\ch{Te-Te}}^{FZ}(q)$ gives a dominant contribution to the total structure factor. One notices that 
$S_{\ch{Te-Te}}^{FZ}(q)$ features a narrow dominant peak centered at around 2.0 \AA$^{-1}$ for x = 0.10 and corresponds to the main peak observed in the total X-ray structure factor (see Fig.[\ref{fig1}]). This peak slightly shifts towards larger $q$ distances with increasing \ch{Na2O} concentration. In addition, we find that the peak occurring at around 1.4 \AA$^{-1}$ exhibits an increasing intensity and shifts towards higher q values as a function of increasing \ch{Na2O} concentration. A similar trend is also observed in the case of $S_{\ch{Te-O}}^{FZ}(q)$ first peak occurring around 1.4 \AA$^{-1}$, thereby indicating that \ch{Te-Te} and \ch{Te-O} correlations are at the origins of the observed FSDP in the total X-ray structure factor (see Fig.[\ref{fig1}]). Finally, by looking at $S_{\ch{O-O}}^{FZ}(q)$, $S_{\ch{Na-O}}^{FZ}(q)$, $S_{\ch{Na-Na}}^{FZ}(q)$ and $S_{\ch{Na-Te}}^{FZ}(q)$, beside typical statistical fluctuations, no significant trends are observed when varying the system composition. 
The overall picture stemming from the reciprocal space analysis, hints towards particular arrangements at intermediate range distances of the glassy network where Te and O play a dominant role.

\subsubsection{Pair distribution function analysis}

\begin{figure}[!h]
	\centering
	\includegraphics[width=0.99\linewidth,keepaspectratio=true]{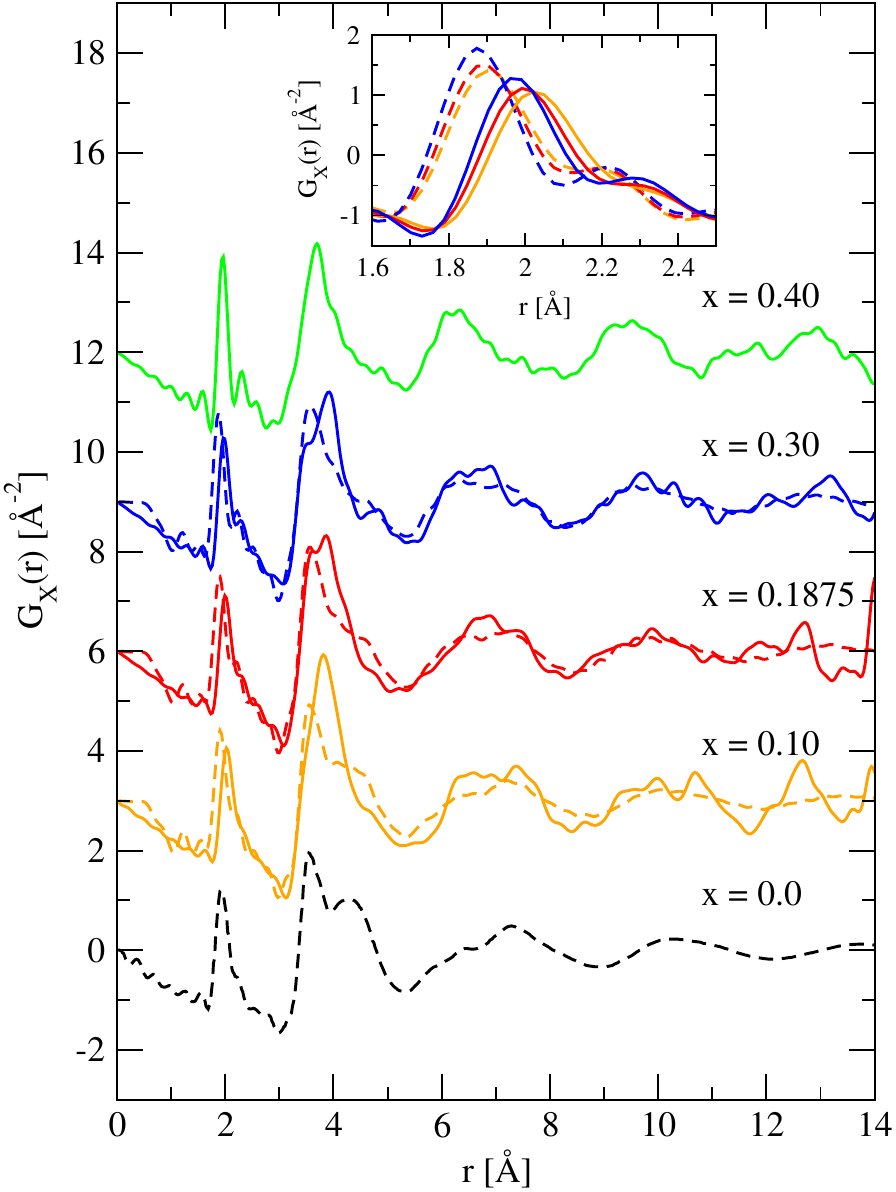}
	\caption{The measured total X-ray pair correlation function G$_{\rm X}$(r) (dashed lines) for the \ch{(TeO2)_{1-x}-(Na_{2}O)_{x}} glasses with x= 0.10, 0.1875, 0.30 and 0.40, compared to the obtained results from FPMD models (solid lines). The PDF of pure \ch{TeO2} taken from Ref.[\citenum{raghvender2022structure}] is also added. The curves are shifted vertically for clarity.}  	
	\label{fig3}
\end{figure}

The computed X-ray PDFs are compared to their experimental counterparts as  shown in Fig.[\ref{fig3}]. Over the entire real space range, we remark that the obtained computed PDFs reproduce the main features of the  measured ones. The first peak, located at around 1.9 {\AA} is well defined, intense and narrow and corresponds to Te-O average bond length. 
We note that the calculated first peak is shifted towards larger $r$ distances by $\approx$ 0.1 {\AA} in comparison to its experimental counterpart. This displacement can be ascribed to the GGA functional's tendency to overestimate bond lengths \cite{hinuma2017comparison, csonka2009assessing, he2014accuracy}. 
Such an overestimation can be corrected by resorting to hybrid exchange and correlation functionals such as PBE0 that showed better performances than PBE functional \cite{kassem2020chemical,matsushita2011comparative,linnera2017ab, raghvender2022structure}. More importantly, the first peak of both experimental and calculated PDFs moves towards smaller $r$, together with an increase of the peak maximum, as increasing \ch{Na2O} concentration.

At slightly larger distances, we observe a collection of low intensity peaks at $\approx$ 2.1–2.4 {\AA}, a distance range consistent with \ch{Na–O} bond lengths found in many crystalline oxide compounds \cite{gagne2016bond}. The intensity of these peaks increases as with increasing \ch{Na2O} concentration. 
The next two intense peaks located at 3.5 {\AA} and 4.3 {\AA} are broader.
In the case of pure \ch{TeO2}, these peaks were assigned to Te-Te distances in 
Te-O-Te bridges and Te${--}$Te occurring at spatial proximity, but not sharing chemical bonds, respectively \cite{raghvender2022structure}. We find that the peak located at around 4.3 {\AA}, transforms into a shoulder and loses its intensity, when increasing the \ch{Na2O} concentration. 
At larger distances, both calculated and experimental PDFs show broad and damped peaks reflecting the absence of long range order in the glasses. 

Despite an overall good agreement between the calculated and the experimental PDFs, this $r$ range around 4.3 {\AA} in the PDF shows the largest discrepancies essentially due to the lack of a perfect description of the Te-Te correlations. 
In fact, it was recently shown in Ref.[\citenum{raghvender2022structure}] that an accurate FPMD modeling of Te-Te subnetwork, especially in pure \ch{TeO2}, is a challenging task and requires the use of hybrid exchange and correlation functionals. 
In this work, we limited our investigation to standard GGA based FPMD for two main reasons: first, although this level of theory cannot clearly separate the closest Te-Te distances involved or not in Te-O-Te bridges, it achieves an overall good reproduction of the network connectivity as reflected by the good match between experimental and calculated PDFs at distances larger than 5 {\AA}.
Second, the addition of \ch{Na2O} to the pure \ch{TeO2} glass leads to substantial structural changes where the Te-Te subnetwork is dramatically altered, as it is demonstrated by the vanishing of the peak at 4.3 {\AA}. 
As such, the adopted GGA level of theory should thereby be sufficient for the analysis put forward in this study. 

\begin{figure}[!h]
	\centering
	\includegraphics[width=0.8\linewidth,keepaspectratio=true]{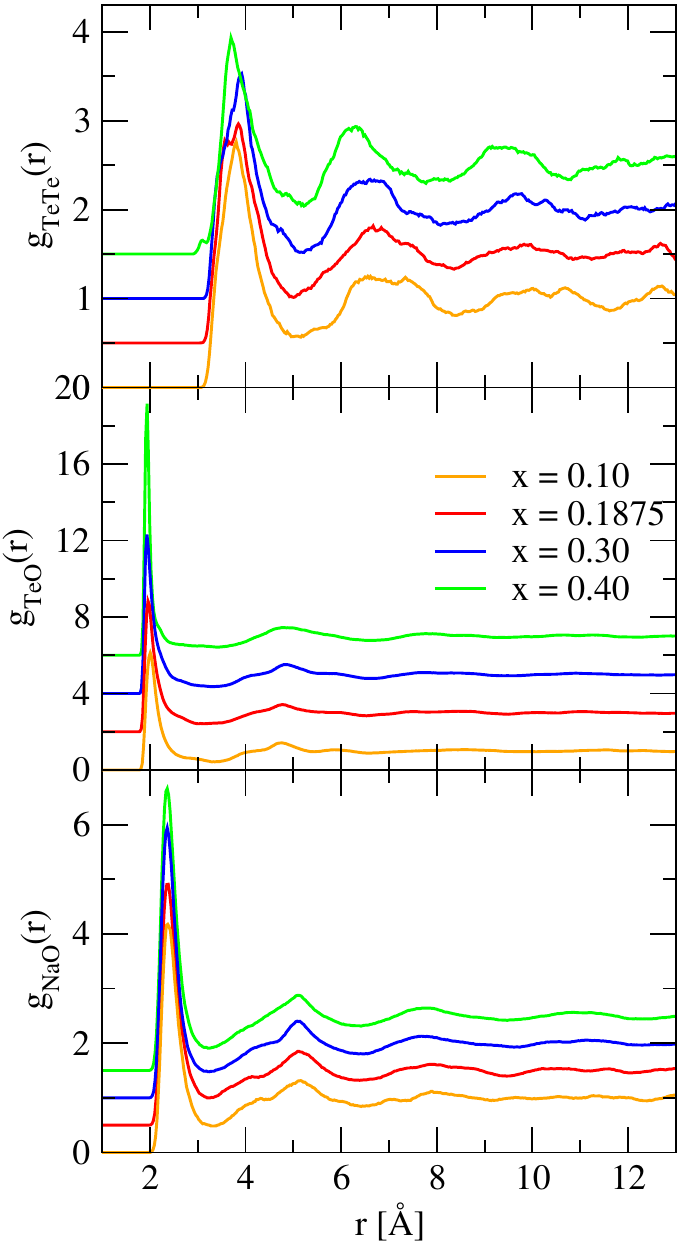}
	\caption{Te-Te, Te-O and Na-O partial pair distribution functions g$_{\alpha\beta}$ for \ch{(TeO2)_{1-x}-(Na_{2}O)_{x}} systems with x= 0.10, 0.1875, 0.30 and 0.40 obtained from FPMD. The curves are shifted vertically for clarity.}  
	
	\label{fig4}
\end{figure} 

\begin{figure}[!h]
	\centering
	\includegraphics[width=0.99\linewidth,keepaspectratio=true]{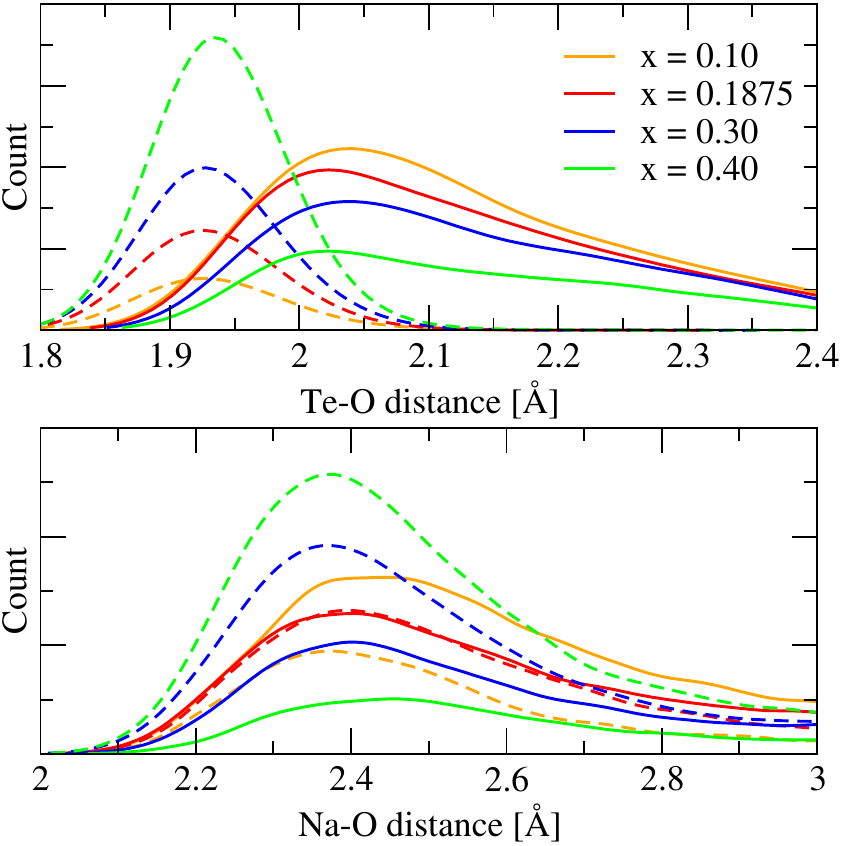}
	\caption{The distribution of Te-O (top panel) and Na-O (bottom panel) distances, broken down into contributions of non-bridging oxygens (shown by dashed lines) and bridging oxygens (represented by solid lines) for \ch{(TeO2)_{1-x}-(Na_{2}O)_{x}} systems with x= 0.10, 0.1875, 0.30 and 0.40 obtained from FPMD. All the distributions are normalized with respect to the number of Te and Na atoms in each system. 
    }  
	\label{fig5}
\end{figure} 

     \begin{table}[h!]
 	\begin{center}
        \scalebox{0.9}{
 		\begin{tabular}{l l l l l}
 			\toprule 
 			                & x = 0.10                & x = 0.1875             & x = 0.30                & x = 0.40\\
		\midrule 
              \ch{Te-O}\, [\AA]& 2.01                   & 1.96                  & 1.94                   & 1.94 \\
              \ch{Te-Te} [\AA]& 3.81                    & 3.85                  & 3.92                   & 3.69 \\
              \ch{Na-O}\, [\AA]& 2.37                    & 2.36                   & 2.36                    & 2.36  \\
 			\midrule
 			
 		\end{tabular}}
 		\caption{The shortest atomic distance of the various pair types obtained from the position of the first peak in the partial PDFs for of \ch{(TeO2)_{1-x}-(Na_{2}O)_{x}} systems with x= 0.10, 0.1875, 0.30 and 0.40 obtained from FPMD.} 		
 		\label{table1}
 	\end{center}
 \end{table}

The calculated Te-O, Na-O and Te-Te partial PDFs are displayed in Fig.[\ref{fig4}]. The Te-O, Na-O and Te-Te shortest distances extracted from the partial PDFs are gathered in Table \ref{table1}. The other partial PDFs are provided in Fig.S1 in the supplementary materials.

Regarding Te-O correlation, we observe that calculated shortest atomic distance decreases from 2.01 {\AA} to 1.94 {\AA} as increasing the \ch{Na2O} content. This trend corresponds to the observed shift in both calculated and experimental 
total PDFs discussed earlier. In order to understand the origins of this shift, we partitioned the Te-O bond distribution into the contributions of Te linked to a bridging oxygen or to a non-bridging oxygen as plotted in Fig.[\ref{fig5}] (top panel). We observe that as a function of increasing the modifier oxide concentration, the intensity of the Te-BO bond distribution reduces while that of Te-NBO bonds increases indicating that Te-BO progressively transform into Te-NBO units.
Coming to the slight shift towards short distances, as the Te-NBO bonds show higher contributions when Na concentration increases, with an average distance of 0.15 {\AA} shorter than Te-BO bonds (see Fig.[\ref{fig5}]), it leads to an overall decrease of the average Te-O bond length. 

Considering the first peak of \ch{g_{NaO}}(r), we find that its position does not significantly change as a function of the modifier concentration. The decomposition of the Na-O distances into Na-BO and Na-NBO is plotted in Fig.[\ref{fig5}] (bottom panel). We find that the population of the Na-BO decreases and transforms into Na-NBO with increasing \ch{Na2O} concentration. However, the difference of the ionic Na-BO and Na-NBO bond lengths is much less marked than what has been observed in the case of Te-O. In addition, this difference compensates, leading to an almost constant average Na-O bond length.

Interestingly, we find that aside from the Te-BO-Te and Te-NBO-Na  bonds, there exists a third form of oxygen bonding identified as BO-Na linkages. These linkages include oxygen being attached to two Te atoms and one Na ion, resulting in BO being triply bonded. Similar result was reported in silicates based glasses \cite{nesbitt2015experimental, du2004medium, cormack2003sodium, mead2006molecular, tilocca2006structural, kargl2006formation, machacek2010md}, however, to the best of our knowledge, there have been no reports in the literature confirming the existence of BO-Na links in the \ch{(TeO2)_{1-x}-(Na_{2}O)_{x}} binary glasses.

Focusing on \ch{g_{TeTe}}(r), one essentially notices that the peaks corresponding to the second and third coordination shells move toward smaller r values with increasing \ch{Na2O} concentration and become well defined. This evolution is consistent with the evolution of the FSDP position and intensity described above. 

\subsubsection{Coordination numbers}

\begin{figure}[!h]
	\centering
	\includegraphics[width=0.99\linewidth,keepaspectratio=true]{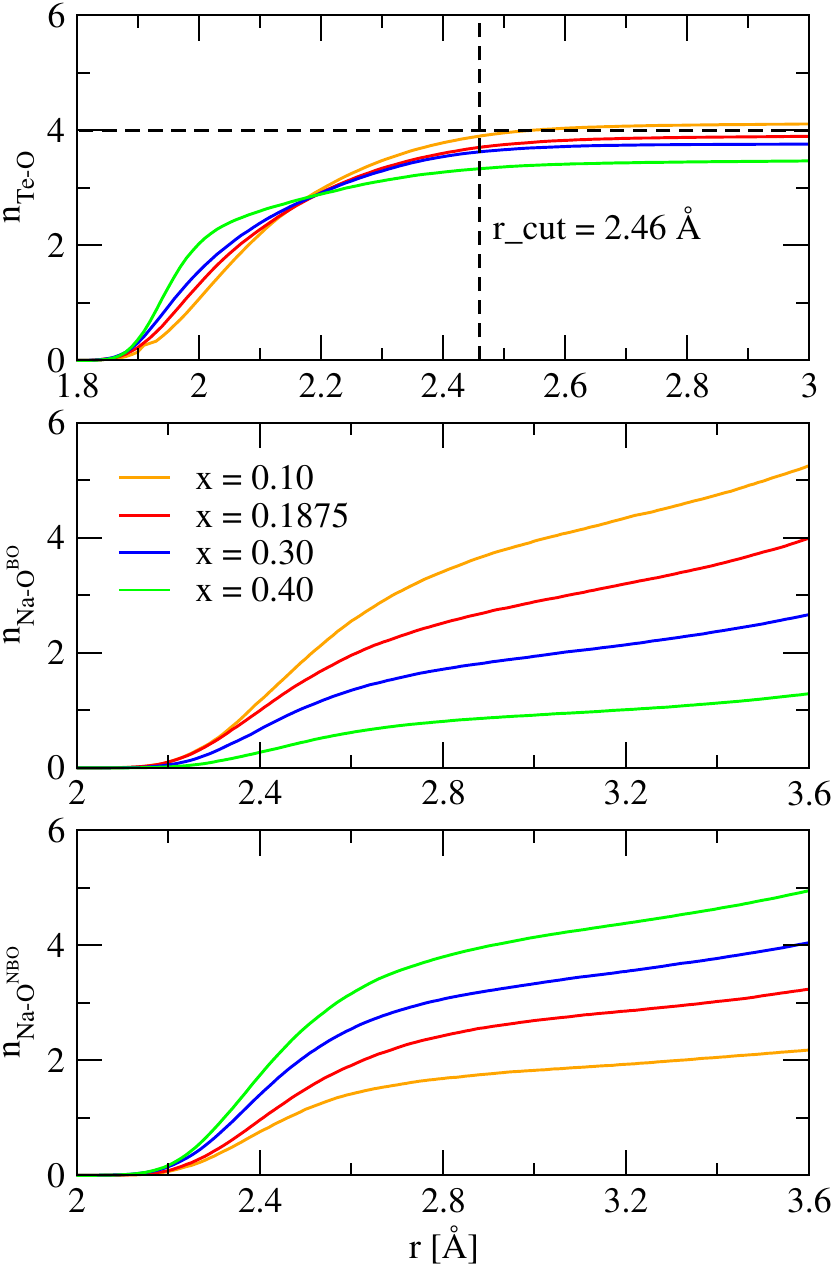}	
 \caption{The running coordination numbers \ch{n_{Te-O}(r)} (top panel) and \ch{n_{Na-O}(r)} as obtained from  MLWF procedure for \ch{(TeO2)_{1-x}-(Na_{2}O)_{x}} glasses with x= 0.10, 0.1875, 0.30 and 0.40 obtained from FPMD at T = 300 K. The coordination number of \ch{Na} is splitted into the one corresponding to bridging BO (middle panel) and non-bridging NBO (bottom panel) oxygens.}  
	\label{fig6}
\end{figure}

The running coordination number of Te (\ch{n_{Te-O}(r)}) and Na (\ch{n_{Na-O}(r)}) atoms calculated based on the MLWF formalism \cite{resta1999electron,marzari2012maximally} are show in Fig.[\ref{fig6}]. 

In the case of Te, we observe that the running coordination number steeply increases at small r values before reaching a plateau region around 2.5 {\AA}. Note that such a plateau cannot be obtained using the conventional method of the partial PDF integration, (i.e. without the use of the MLWF formalism) as shown in Fig.S2 in the supplementary materials. The initial increase in the \ch{n_{Te-O}(r)} is more pronounced with larger \ch{Na2O} content due to the larger fraction of Te-NBO.
In the case of Na, the running coordination numbers are broken down into contributions of BOs and NBOs. They both show a similar evolution compared to Te, yet without reaching a plateau at large r values. This can be ascribed to the ionic nature of Na-O bond which allows for a large variety of Na local environments. 

In practice, obtaining atomic coordination numbers requires a proper definition of a cutoff distance. Fig.S3 in supplementary material displays the distribution of Wannier centers around \ch{Te}, \ch{O} and \ch{Na} atoms computed for all the studied systems. Regardless of the system composition and the chemical species, the first peak in the distribution of $w_{n}$(r) is a result of \ch{W^{LP}} at the vicinity of the central atom. The second peak is attributed to bonding Wannier \ch{W^{B}}, and the last peak is due to \ch{W^{LP}} at the vicinity of the first neighboring atoms. In the case of Te-O bonds, the values of the \ch{Te-W^{B}} and \ch{O-W^{B}} second minimum positions in Fig.S3 can be used to estimate a \ch{Te-O} cutoff distance. When added together, these values offer a good estimation of \ch{Te-O} cutoff with a value of 2.46 {\AA}, in agreement with values reported in the literature in the case of pure \ch{TeO2} glass \cite{raghvender2022structure}.
As for \ch{Na-O} distances, the ionic nature of the Na-O bond makes the identification of the bonding Wannier difficult as these centers are very close to O atoms preventing the occurrence of a clear \ch{Na-W^{B}} minimum positions as shown in Fig.S3. Therefore, we used 2.93 {\AA}, as a cutoff distance for the identification of \ch{Na-O} bonds. This value corresponds to the first minimum in the  partial Na-O pair correlation function (see Fig.[\ref{fig4}]) and to the largest \ch{Na-O} bond length found in the crystalline phase \ch{Na2Te4O9} \cite{tagg1994crystal}, whose composition almost corresponds to that of the glass at x = 0.1875.
Table \ref{table2} summarizes the obtained coordination numbers for Te, Na and O atoms. 

\begin{table*}[!htbp]
\centering
\scalebox{0.65}{
\begin{tabular}{c c c c c}  
\toprule
\hline
	      x             & 0.10           &   0.1875       & 0.30           & 0.40\\
\hline   
	\midrule
	  \ch{n_{Te-O}}     & 3.90$\pm$0.03  & 3.70$\pm$0.04  & 3.62$\pm$0.04  & 3.33$\pm$0.03  \\
    \midrule
    \ch{n_{Na-O}}     & 5.57$\pm$0.10  (3.77$^{\rm{BO}}$+1.80$^{\rm{NBO}}$)  &  5.38$\pm$0.08 (2.76$^{\rm{BO}}$+2.62$^{\rm{NBO}}$)  & 5.10$\pm$0.06 (1.86$^{\rm{BO}}$+3.24$^{\rm{NBO}}$)  & 4.98$\pm$0.04 (0.85$^{\rm{BO}}$+4.13$^{\rm{NBO}}$) \\
    \midrule
    \ch{n_{O-Te}}     & 1.85$\pm$0.02  &  1.66$\pm$0.02 & 1.49$\pm$0.02  &  1.22$\pm$0.01  \\
    \midrule
    \ch{n_{O-Na}}    & 0.58$\pm$0.02   &  1.11$\pm$0.03 & 1.80$\pm$0.04  &  2.49$\pm$0.01  \\
    \midrule
    \ch{n_{O}}=\ch{n_{O-Te}}+ \ch{n_{O-Na}}             & 2.43$\pm$0.02  & 2.77$\pm$0.02  & 3.29$\pm$0.03  &   3.71$\pm$0.02   \\
    \midrule
\hline
\end{tabular}}
\caption{The calculated \ch{n_{$\alpha \beta$}} coordination number and its corresponding error bars for the \ch{(TeO2)_{1-x}-(Na_{2}O)_{x}} glasses with x= 0.10, 0.1875, 0.30 and 0.40 obtained from FPMD. 
}
\label{table2}
\end{table*}

\begin{figure}[h!]
	\centering
	\includegraphics[width=0.99\linewidth,keepaspectratio=true]{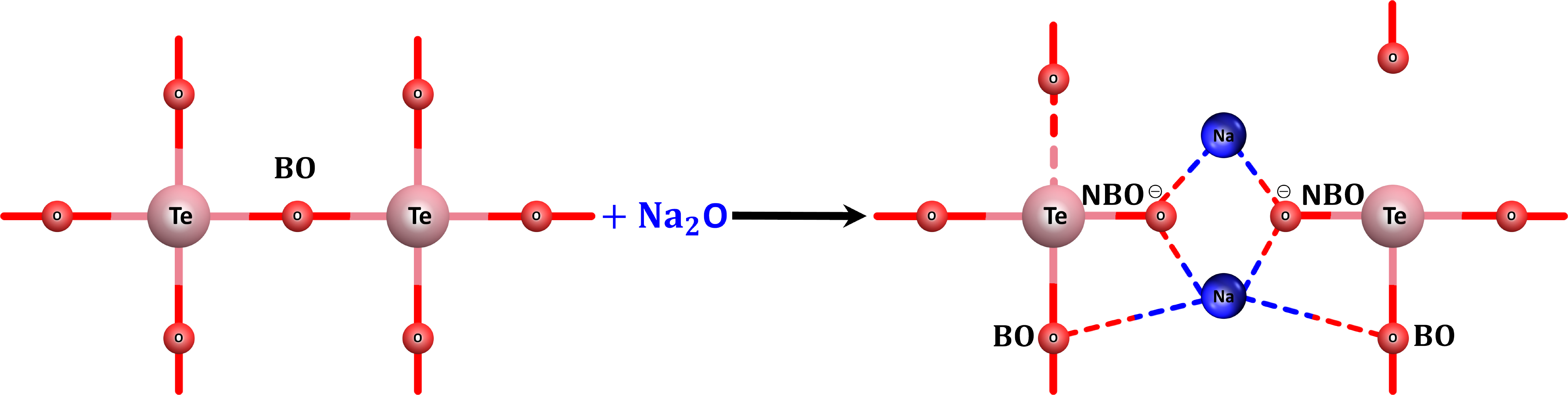}
	\caption{Schematic representation of the \ch{Te-O-Te} network structurally depolymerized by adding \ch{Na2O} modifier.}  
	\label{fig7}
\end{figure}

By looking at the coordination number of Te atoms, \ch{n_{Te-O}} shows a substantial reduction from 3.90 to 3.33 going from x = 0.10 to 0.40. This indicates a strong structural depolymerization of the \ch{Te-O} network which goes along with the transformation of \ch{TeO4} units into \ch{TeO3} units previously discussed. This is schematically illustrated in Fig.[\ref{fig7}].
This depolymerization already occurs at concentrations as low as x = 0.10 as reflected by the   \ch{n_{Te}} value of 3.90 for x = 0.10 as compared to 3.96 obtained for the pure \ch{TeO2} glass \cite{raghvender2022structure}.

Regarding sodium atoms, we find that the \ch{Na} coordination number moderately reduces from 5.57 for x = 0.10 to 4.98 for x = 0.40.  
This trend is in agreement with experimental results where it was reported that \ch{n_{Na}} decreases from 5.2 for x = 0.10 to 4.6 for x = 0.19 when using a cutoff distance of 2.37 {\AA} \cite{barney2015alkali}. 
In addition, similar results were obtained in sodium silicate glasses at x=0.425 where the coordination number was found to be 4.98 when considering a cutoff distance of 2.45 {\AA} \cite{hannon2021structure}. 
Interestingly, this moderate evolution of the \ch{Na} coordination number is the
result of a large but opposite evolution of the BO and NBO contributions, \ch{Na-BO} and \ch{Na-NBO}, to the total \ch{Na} coordination number. This kind of evolution has been reported in the case of sodium silicate glass. In the particular case of x = 0.40, Du and Cormack \cite{du2004medium} reported calculated \ch{Na-O} coordination number of 5.1 where Na is bonded to 3.9 NBOs and 1.2 BOs, and Hannon \textit{et al.}\cite{hannon2021structure} reported measured \ch{Na-O} coordination number of 4.8. 

Coming to oxygen atoms, as \ch{Na2O} is added, \ch{n_{O-Te}} decreases, while the  \ch{n_{O-Na}} increases leading to an overall increase in the average coordination number of O when increasing the \ch{Na2O} content. 

\subsubsection{Atomic local environments and short range disorder}

\begin{table}[htbp!]
 	\begin{center}
 		\begin{tabular}{l l l l l l}
                \toprule
 			\toprule 
 			${\alpha}$ & $Q_{m}^{n}$ (\%) & x=0.10 & x=0.1875 & x=0.30 & x=0.40\\
 			\midrule 
             Te   & $Q_{3}\, $  & 27.9 & 38.3 & 43.8 & 66.8 \\
                & $Q_{3}^{0}\, $  & - & - & - & 22.8\\
 			& $Q_{3}^{1}\, $  & - & 7.3&19.2 & 33.5 \\
 			& $Q_{3}^{2}\, $ & 13.7 & 24.5 & 20.4& 10.0 \\
 			& $Q_{3}^{3}\, $  &11.4 & 6.2 & -& -  \\
            \cmidrule(lr){2-6} 
              &   $Q_{4}\, $  & 53.9 & 52.2 & 49.6 & 30.3 \\
 			& $Q_{4}^{2}\, $  & - & - & 14.6& 18.0 \\
              &  $Q_{4}^{3}\, $  & 20.7 & 30.0 & 27.7& 11.0 \\
 			& $Q_{4}^{4}\, $ & 31.9 & 18.8 & 7.2 & - \\
                \cmidrule(lr){2-6} 
              &  $Q_{5}\, $  & 18.0 & 9.1 & 6.4 & - \\
 			& $Q_{5}^{4}\, $  & - & - & -& - \\
 			& $Q_{5}^{5}\, $  & 15.8 & 6.3 & -& -\\
 			\midrule
     		\midrule
          Na   &   $Q_{4}\, $  & 8.3 & 11.9 & 17.4 & 21.2 \\
                \cmidrule(lr){2-6} 
              &  $Q_{5}\, $  & 38.3 & 43.4 & 53.4 & 56.8 \\
 			\cmidrule(lr){2-6}
               &  $Q_{6}\, $  & 41.1 & 37.0 & 25.7 & 19.5 \\
 			\cmidrule(lr){2-6}
          &  $Q_{7}\, $  & 10.5 & 6.8 & 2.2 & - \\
                 \midrule 
                  \midrule 
 		\end{tabular}
 		\caption{Percentage of $Q_{m}^{n}$ computed based on the MLWF formalism and  r$_{\text{cut}}$ = 2.46 {\AA} for \ch{Te}, and r$_{\text{cut}}$ = 2.93 {\AA} for \ch{Na} for \ch{(TeO2)_{1-x}-(Na_{2}O)_{x}} systems with x= 0.10, 0.1875, 0.30 and 0.40 obtained from FPMD. Values of $Q_{m}^{n}$ less than $\sim$2\% are not reported.} 		
 		\label{table3}
 	\end{center}
 \end{table}

We analyse the atomic local environment in terms of $Q_{m}^{n}$ structural units, where $m$ is the number of the nearest neighbors and $n$ is the number of BO\cite{pietrucci2008teo, gulenko2014structural}. The results of this decomposition are displayed in Table \ref{table3} were $Q_{m}^{n}$ fractions lower than $\sim$ 2 \% are not reported.

For x= 0.1, the Te local environment shows a dominant fraction of 4-fold units (53.89 \%) followed by a large fraction of 3-fold coordination environments (27.89 \%) and a non negligible fraction of 5-fold coordination number (18.02 \%). As the \ch{Na2O} content increases, the population of \ch{TeO4} and \ch{TeO5} units reduce while we observe a substantial increase of the \ch{TeO3} population where 
at x= 0.4, the 3-fold Te units become dominant with a fraction of 66.83 \%. 

As increasing the fraction of \ch{Na2O}, one notices a decrease in the number of fully connected \ch{TeO3} ($Q_{3}^{3}$) units, as well as the increase in the number of $Q_{3}^{1}$. In the particular case of x = 0.40, we find a significant fraction of isolated \ch{TeO3} ($Q_{3}^{0}$) units. As for the $Q_{3}^{2}$ units, they reach a maximum value at a modifier oxide concentrations of x = 0.1875 and 0.30, whereas they remain around 10-13\% for the lowest and the highest modifier concentration. 
In the case of 4-fold Te atoms, a similar analysis to that of the 3-fold Te atoms applies. Specifically, when increasing the modifier concentration, the fraction of $Q_{4}^{2}$ units increases, and that of $Q_{4}^{4}$ decreases, while the fraction of $Q_{4}^{3}$ shows a maximum around x = 0.1875 and 0.30.
When considering over-coordinated Te polyhedra, it is seen that the  $Q_{5}^{5}$ population rapidly decreases as a function of increasing the \ch{Na2O} content. 
These observations again demonstrate the contribution of \ch{Na2O} units to the structural depolymerization of \ch{TeO2} network by replacing Te-BO-Te bridges with Te-NBO-Na bridges. 

Considering \ch{Na} atoms, we find that they show substantial fractions of 5-fold and 6-fold Na units together with a small fraction of \ch{NaO4} and \ch{NaO7} units at x = 0.1. We notice that the population of 4-fold and 5-fold units increase as the \ch{Na2O} content increases, while the opposite is observed for the population of 6-fold and 7-fold units. This trend can be attributed to the decrease in the population of Na-BO bonds while exhibiting a substantial increase in the population of Na-NBO bonds as the concentration of \ch{Na2O} increases.

\subsubsection{Bond angle distributions}

\begin{figure}[!h]
	\centering
	\includegraphics[width=0.99\linewidth,keepaspectratio=true]{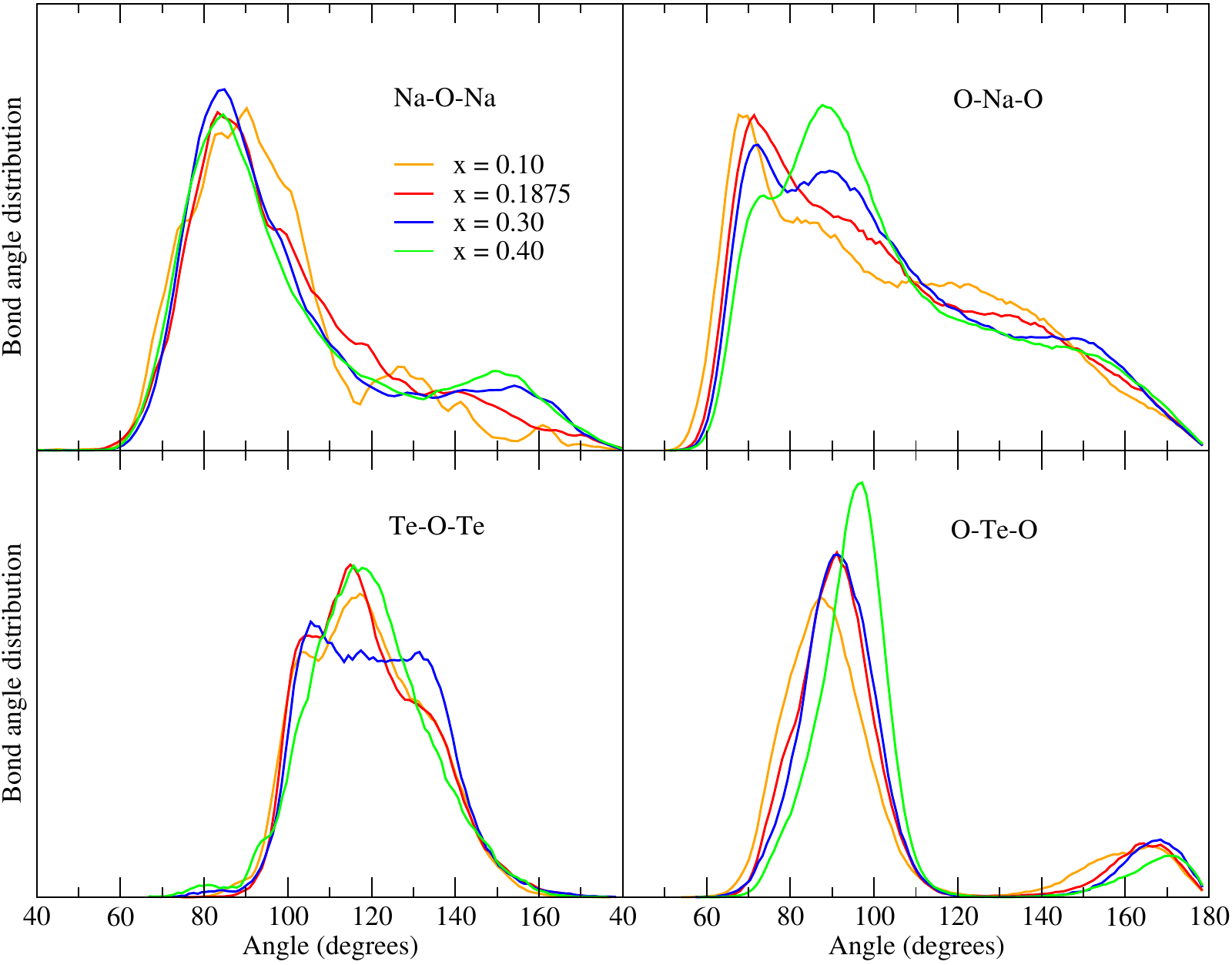}
	\caption{The bond angle distribution around Te , O and Na atoms for \ch{(TeO2)_{1-x}-(Na_{2}O)_{x}} glasses with x= 0.10, 0.1875, 0.30 and 0.40 obtained from FPMD at T = 300 K. We use a cutoff distance similar to the one used to calculate the coordination number to compute the bond angle distributions.}  
	\label{fig8}
\end{figure}

The bond angle distributions (BAD) around Te, O, and Na atoms are shown in Fig.[\ref{fig8}]. 
Focusing on O-Te-O BAD, it features a first intense peak at around 90$^{\circ}$ and a second less intense one at 165$^{\circ}$. These angles are typical of those found in pure \ch{TeO2} glass, where it has been reported that the peak between 70$^{\circ}$ and 110$^{\circ}$ corresponds to O-Te-O equatorial linkages while the one between 150$^{\circ}$ and 180$^{\circ}$ corresponds to axial O-Te-O linkages \cite{gulenko2014atomistic}. 

As \ch{Na2O} concentration increases, this first peak intensity slightly increases and its position shifts from 87.12$^{\circ}$ for x = 0.10 to 91.13$^{\circ}$, 91.13$^{\circ}$, and 97.16$^{\circ}$ for x = 0.1875, 0.30, and 0.40, respectively. Similarly, the second peak becomes narrower with an average position shifting to larger angles. In order to relate these evolutions to the modification of the network connectivity, we broke down the O-Te-O bond angle distributions into the contributions from atoms within \ch{TeO3}, \ch{TeO4} and \ch{TeO5} units. Additionally, we also partitionned the O-Te-O BAD depending on the nature of the O atoms: BO and NBO. 
The results are shown in Fig.S4 and Fig.S5 in the supplementary material.
We notice that as the modifier oxide concentration rises, the contribution from atoms belonging to the 4- and 5-fold Te units decreases while those belonging to 3-fold Te units increases and become dominant at x = 0.40. This trend corresponds to 
transformations of the 4- and 5-fold units into 3-fold units leading to a higher peak intensity and narrower distribution. Furthermore, as 3-fold Te units are characterized by larger O-Te-O angles than those found in 4- and 5-fold Te units, the main distribution shifts towards larger angles (Fig.\ref{fig8}). Moreover, the angles between 150$^{\circ}$ and 180$^{\circ}$ are found to be essentially due  to 4-fold Te units. In addition, the BO-Te-BO and NBO-Te-BO distributions are found to have a substantial contribution to the first O-Te-O peak at x = 0.10. When Na concentration increases, the breaking of Te-O-Te bridges causes an increase in the NBO population, which consequently leads to an increase in the NBO-Te-NBO and NBO-Te-BO distributions and becomes dominant at x = 0.40, with a shift of the central peak position. The second less intense angle at 170$^{\circ}$ is observed to be related to the NBO-Te-BO and the BO-Te-BO distributions and due to the aforementioned reason,  its population decreases as the concentration of \ch{Na2O} increase. 

Coming to Te-O-Te BAD, it is indicative of the interconnections between Te-based structural units. This distribution is shown in Fig.[\ref{fig8}] and its decomposition based the coordination number of Te atoms is plotted in Fig.S6. 
We find that for all x values, the Te-O-Te distribution is centered at about 120$^{\circ}$ which corresponds to typical angles found in pure \ch{TeO2} glass and in the \ch{TeO2} crystalline polymorphs \cite{gulenko2014atomistic}. Nevertheless, the distribution is broad with several distinguishable contributions. 
For x = 0.10, we find that the peak at $\sim104^{\circ}$ is mainly due to the contribution of TeO$_{4}$-O-TeO$_{4}$ and TeO$_{4}$-O-TeO$_{5}$ environments, while the other two peaks at $\sim117.2^{\circ}$ and $\sim130^{\circ}$ are mostly due to the TeO$_{3}$-O-TeO$_{4}$, TeO$_{4}$-O-TeO$_{4}$ environments with a small contribution of TeO$_{4}$-O-TeO$_{5}$ environments. When modifier oxide is added, the TeO$_{3}$-O-TeO$_{4}$ and TeO$_{4}$-O-TeO$_{4}$ environments lead to the peaks observed at about $\sim 100^{\circ}$ and $\sim 120^{\circ}$ and give rise to the first and second peaks in the total  Te-O-Te bond angle distribution for x $<$ 0.30. 
In the case of x = 0.30, one can notice that the TeO$_{4}$-O-TeO$_{4}$ environments show a high contribution to the peak around $\sim 100^{\circ}$, while TeO$_{3}$-O-TeO$_{4}$ lead to the peaks observed near $\sim 117^{\circ}$ and $\sim 130^{\circ}$ which results in the absence of clear maxima in the Te-O-Te angle distribution for angles larger than $110.0^{\circ}$. Finally, for x = 0.40, the main contribution comes from TeO$_{3}$-O-TeO$_{4}$ leading to a single peak centered at $\sim 117^{\circ}$. 

Finally, coming to \ch{O-Na-O} and \ch{Na-O-Na}, they exhibit much broader BADs than the BADs related to Te atoms. As already stated above, this can be ascribed to the ionic nature of Na-O bond which allows for a large variety of Na local environments. 

\subsection{Dynamical properties}

\subsubsection{Na diffusion and ionic conductivity}

The calculated MSD of Te, O and Na atoms as a function of time are provided in Fig.S7 in supplementary materials. We remark an increase in the MSD value for all the chemical species as a function of increasing temperature. One notices that the MSD of Na and O exhibit a larger increase compared to Te as the temperature increases. 
For all \ch{Na2O} concentrations, the Te atoms are the less mobile chemical species, with a maximum MSD value at 30 ps and T = 1400 K $<$ 6 \AA$^{2}$ . This result supports that the systems are still in the solid phase, where the Te maintains the glassy network, while O and Na are more mobile \cite{pham2023unveiling}.
At T = 900 K, while all chemical species exhibit a very low MSD value ($<$ 3 \AA$^{2}$), Na and O atoms show slightly higher mobility compare to Te atoms, which increases as a function of increasing the \ch{Na2O} content. At T = 1100 K, O atoms have MSD values higher than the Na ions for concentration x $<$ 0.30, while for x $\ge$ 0.30 the Na diffusion become dominant. This behavior persists at higher temperatures and indicates that \ch{(TeO2)_{1-x}-(Na_{2}O)_{x}} glasses behave differently at lower sodium contents (x $<$ 0.30), where Na diffusion is limited (even at high temperature). This trend is potentially due to the encapsulation of Na atoms in cage-like environments formed by over-coordinated sodium units (7-fold units), as stated previously and illustrated in Table \ref{table3}. Compositions with x $\ge$ 0.30 maintain consistently higher Na ion mobility, with MSD values increasing steadily from T = 1100 K to 1400 K. At T = 1400 K, compositions with x $\ge$ 0.30 show Na mobility over 30 ps that is twice  larger than the one found at lower concentration. 
Overall, these results reveal that Na diffusion in glassy systems is governed by the glass composition and the temperature.
 
\begin{figure}[!h]
	\centering
	\includegraphics[width=0.99\linewidth,keepaspectratio=true]{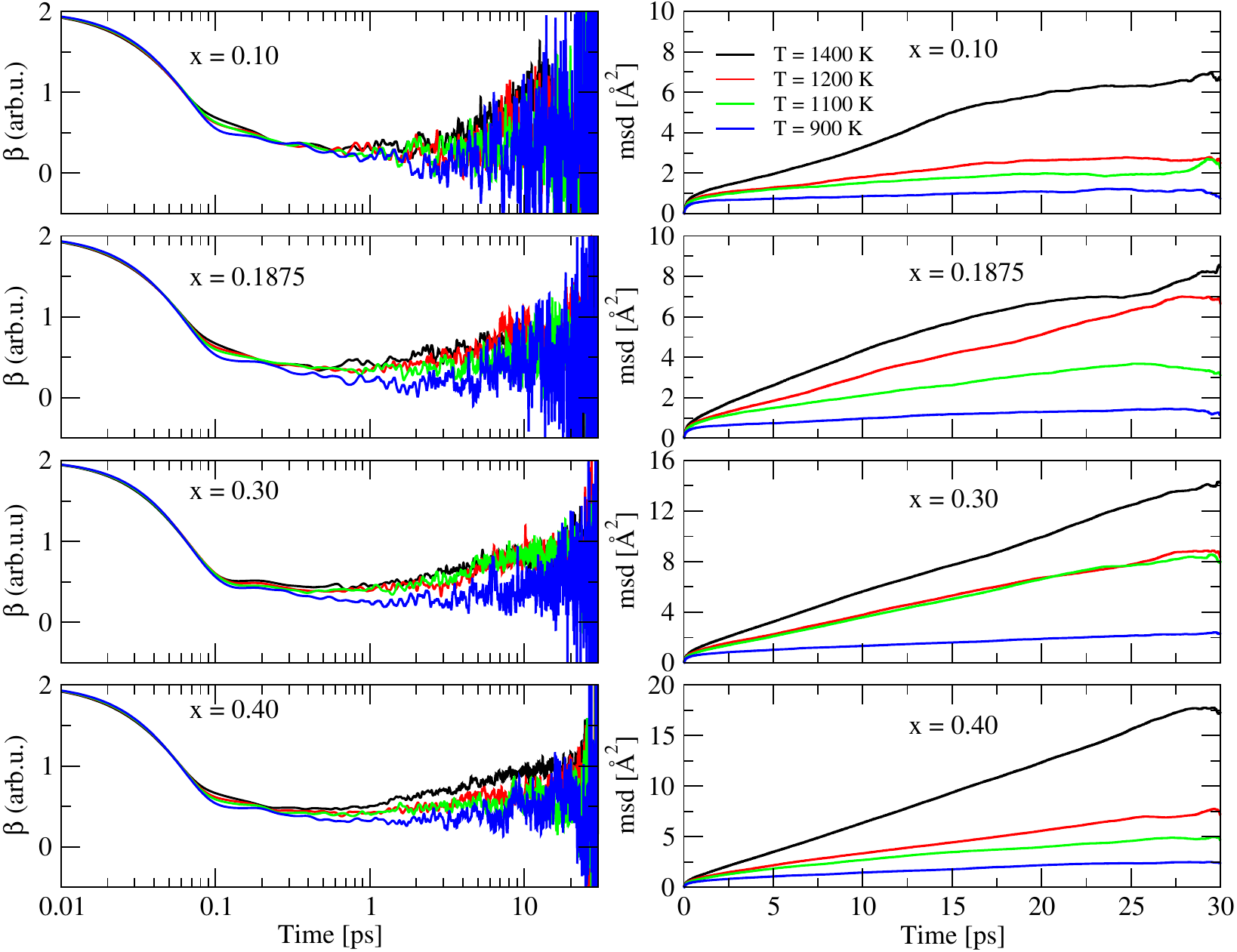}
	\caption{Right panels: plot of MSD $vs.$ time for Na in
 \ch{(TeO2)_{1-x}-(Na_{2}O)_{x}} with x = 0.10, 0.1875, 0.30 and 0.40 glasses at temperatures comprised between T = 900 K and 1400 K. Left panels: corresponding values of $\beta$ $vs.$ time.}  
       \label{fig10}
\end{figure} 

To evaluate the long-range mobility of Na-ions, we analyzed the log-log plot of the mean square displacement (MSD) and $\beta$ (Eq.\ref{d2}) as a function of time (t) at the considered temperatures as shown in Fig.[\ref{fig10}].
Three different regimes of ion motion can be distinguished. The initial stage reveals a ballistic regime, where MSD varies quadratically with time ($\beta$=2). As time increases above 0.1 ps, the system enters a caging regime, where Na$^+$ cations are confined by their local environments. This regime is characterized by $\beta$ values lower than 1 and persists for approximately 10 ps \cite{weeks2002subdiffusion,rips1994cage,pham2023unveiling}. 
Following this phase, Na$^+$ cations enter the diffusive regime, where they travel along diffusion pathways, and the MSD increases linearly with time. This transition marks the onset of long-range diffusion and is characterized by a $\beta$ value approaching 1. The rate at which it reaches this value accelerates with increasing temperature; however, it exhibits increased statistical noise at extended times. 

\begin{figure}[!h]
	\centering
	\includegraphics[width=0.99\linewidth,keepaspectratio=true]{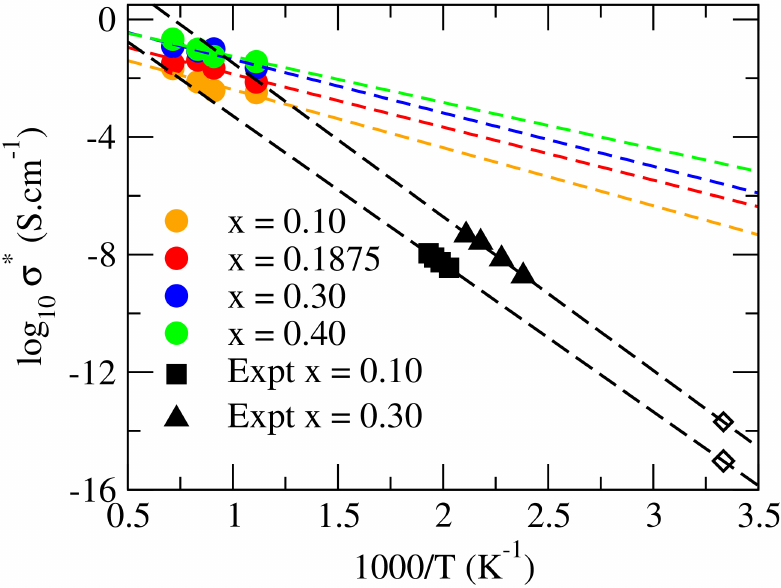}
	\caption{Ionic conductivity of \ch{(TeO2)_{1-x}-(Na_{2}O)_{x}} glasses with x = 0.10, 0.1875, 0.30 and 0.40. Closed black circles correspond to experimental results, while colored ones represent the FPMD data and open diamonds correspond to data at T = 300 K extrapolated by using the linear Arrhenius fit.} 
        \label{fig11}
\end{figure}

The Na self-diffusion coefficients $D_{\text{Na}}^{*}$ evaluated within the diffusion regime and the corresponding ionic conductivities $\sigma_{\text{Na}}^{*}$ are plotted as a function of temperature in Fig.S8 and Fig.[\ref{fig11}], respectively. The measured $\sigma_{\text{Na}}$ at x=0.10 and x=0.30 are also shown in Fig.[\ref{fig11}].

Despite the effect of the limited statistics on the calculated conductivity values, a jump of almost two orders of magnitude of $D_{\text{Na}}^{*}$ and $\sigma_{\text{Na}}^{*}$ is observed between glasses with low (x = 0.10, 0.1875) and high (x = 0.30, 0.40) \ch{Na2O} concentrations at high temperatures. This trend agrees with the change of the extrapolated measured ionic conductivity when going from low to high \ch{Na2O} concentrations. While the high temperature range shows a qualitative agreement between modeling and experiments, the extrapolated calculated ionic conductivity to room temperature show a strong deviation by several orders of magnitude. This discrepancy can be attributed to various aspects including the possible finite-size effects in our models, the high temperature range used to estimate the ionic conductivity (up to T= 1400 K), and the finite simulation time. Since the computed ionic conductivity is extremely low (of the order of $\mu$S/cm), molecular dynamics (MD) runs at elevated temperatures are typically required to induce significant ionic motion over short simulation timescales. Consequently, the extrapolation of the ionic conductivity at room temperature may be biased. Several previous works have highlighted this issue that leads to non-Arrhenius behavior of the calculated conductivity \cite{winter2023simulations,sasaki2025constant,he2018statistical}.

This discrepancy is also confirmed in the activation energy of the diffusion mechanism. The calculated activation energy remains relatively constant around 0.45 eV across all studied compositions, whereas the experimental one is of 0.95 eV and 1.0 eV for x=0.1 and x=0.30, respectively. The high measured experimental values reflects the low ionic conductivity at room temperature 
9.4 10$^{-16}$ and 2.0 10$^{-14}$ S/cm$^2$ for x=0.1 and x=0.30, respectively and fall within the broader range of 0.2 to 1.1 eV of activation energies commonly reported for sodium-conducting glasses. These values of activation energy vary significantly depending on the alkali content and glass network structure \cite{thomas1984electrical, hayashi2019sodium, liu2016promising, hona2023ionic, michalski2019properties, hona2023alkali}

Despite the seemingly inherent statistical errors to FPMD modeling, the qualitative agreement on the evolution of the ionic conductivity as a function of the \ch{Na2O} concentrations suggests that the underlying physical models and simulation parameters employed in the presented study capture the essential transport mechanisms governing the ionic conductivity in these glasses at low temperatures. 

In this context, the observed conductivity jump from low to high \ch{Na2O} concentrations can be correlated to the significant change in the structure of the glass. Specifically, the glass at lowest \ch{Na2O} concentration show a significant fraction of 7-fold ($\approx$ 10\%) Na atoms as well as a large proportion of Na atoms linked to BO atoms. In contrast, at the highest \ch{Na2O} concentrations the glasses
do not contain 7-fold Na atoms and show a dominant proportion of Na linked to NBO atoms.

\subsubsection{Channel Formation and Intermediate Range Order}

\begin{figure}[!h]
	\centering
	\includegraphics[width=0.99\linewidth,keepaspectratio=true]{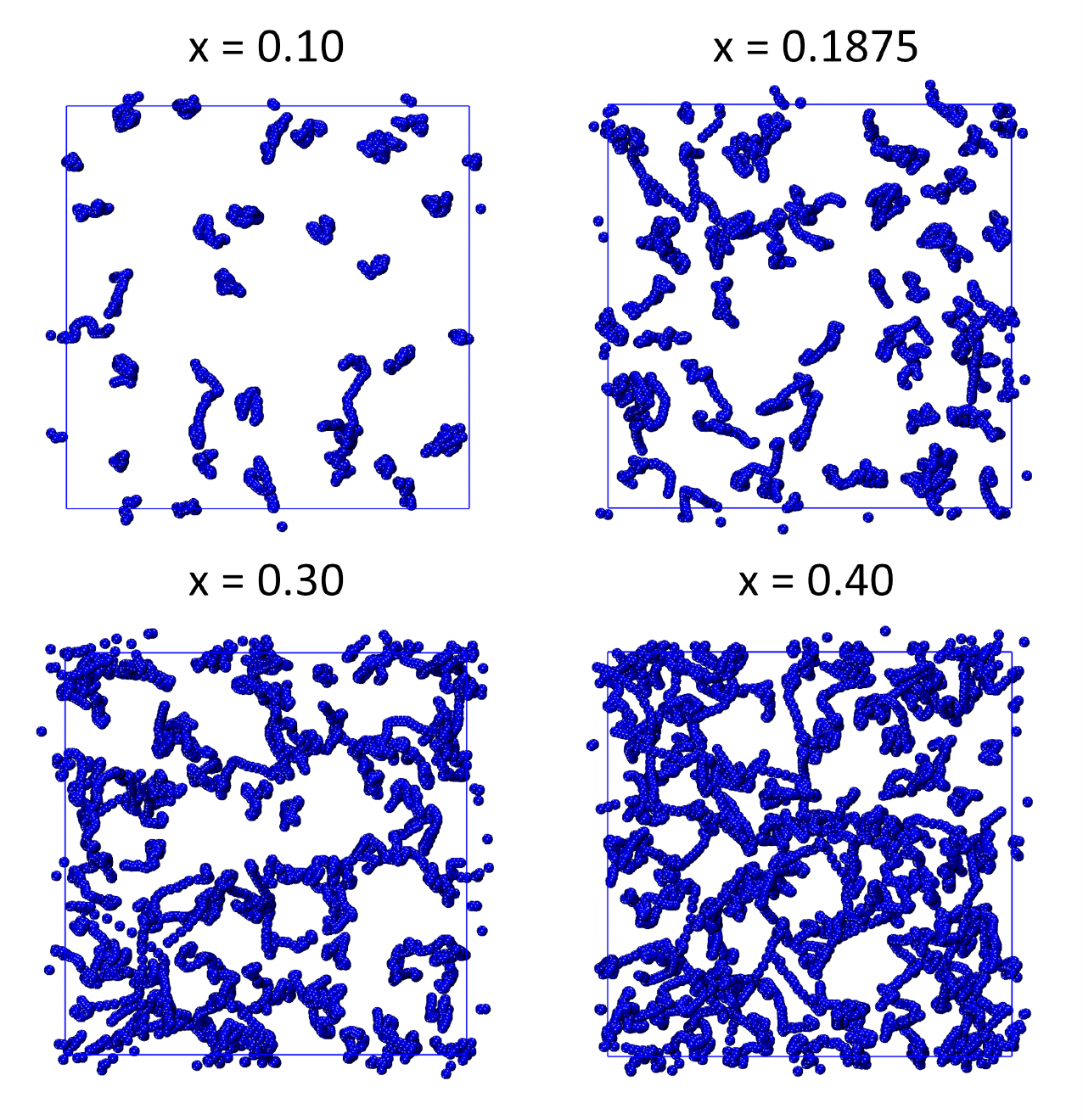}
	\caption{Accumulated trajectories over 30 ps at 1200 K. The blue spheres that are connected to each other represent the Na atoms for \ch{(TeO2)_{1-x}-(Na_{2}O)_{x}} glasses with x = 0.10, 0.1875, 0.30 and 0.40.}  
       \label{fig9}
\end{figure}

The diffusion of sodium ions in \ch{(TeO2)_{1-x}-(Na_{2}O)_{x}} glasses involves a complex process, strongly dependent on the glass structure. Fig.[\ref{fig9}] shows the accumulated trajectory of Na atoms over 30 ps at T = 1200 K for all the studied systems. 
At low concentrations of \ch{Na2O} (x = 0.10, 0.1875) we mainly observe localized Na trajectories. In contrast, at highest concentrations (x = 0.30, x 0.40) we observed the formation of large Na diffusion channels. The presence of these channels correlates well with the high mobility observed at high \ch{Na2O} concentrations, whereas the localized trajectories at low \ch{Na2O} concentrations indicates that Na atoms are confined.

The structural properties of \ch{TeO2}-based glasses, governed by the tellurite network's unique geometry, play a fundamental role in defining these pathways and their evolution. Firstly, sodium acts as a network modifier, disrupting the original tellurite network, which is predominantly composed of \ch{TeO4} trigonal bipyramids. 
Secondly, with increasing \ch{Na2O} content, Na$^+$ cations exhibit a tendency to associate with non-bridging oxygens, leading to the formation of Na-rich regions that progressively evolve into percolating diffusion channels at higher \ch{Na2O} concentrations. This behavior is remindful of the modified random network model proposed by Greaves in silicate glasses \cite{greaves1981local, greaves1994random, ojovan2021modified}. 

A representative long channel for the system with x = 0.30 at T = 1200 K is depicted in Fig.S9. This channel contains seven Na atoms, which diffuse along a long pathway surrounded by Te atoms. Such channels induce a restructuring of the Te sub-lattice at its neighborhood which can be correlated to the well defined FSDP at high \ch{Na2O} concentrations. 

The intensity of the FSDP in the total structure factor increases with the \ch{Na2O} concentration (see Fig. \ref{fig1}). The decomposition of the Faber-Ziman structure factors revealed that this trend is essentially due to an increase of the Te-Te correlations first, then to a less extent to the Te-O correlations. At low concentration the FSDP is barely visible and only strengthens when the \ch{Na2O} concentration exceeds x=0.1875 (see Fig. \ref{fig2}). This observation correlates well with the formation of the percolation channels in the glass. Upon visually inspecting the trajectories, it was found that the Na channels are surrounded by a structured and quasi-stationary Te sub-network (Fig.S9). While a clear assessment of the structure of such channels would require sophisticated methods, such as Voronoi tessellation, a simpler indication can be obtained from the Te-Te pair correlation function. Upon increasing the \ch{Na2O} concentration, all peak positions shift to smaller distances and those occurring beyond 5 {\AA} become sharper. This trend is visible when comparing the Te-Te partial PDFs from models with x=0.10 and x=0.40 models, indicating a denser packing of the Te-sub network and an increased level of structuring at intermediate distances, consistent with the FSDP correlation distances in real space (see Fig. \ref{fig4}). In practice, the high \ch{Na2O} concentration leads to the connection of Na centered cavities, creating percolated channels, and causing the surrounding Te to organize around these channels.
In addition, the Te-sub-network reconstruction is persistent at high temperatures as the FSDP remains visible on the calculated structure factors at high temperatures (see Fig.S10). A similar trend has been described in Na-silicate glasses in a previous work \cite{meyer2004channel}.

Tracking Na atoms over time revealed a correlated push-pull mechanism, illustrated in Fig.S10. It can be seen that a Na atom initiates motion downward, exerting a directional push on adjacent Na$^+$ cations. This action propagates through the channel, causing all participating Na$^+$ cations to move synchronously in the same direction. This collective motion is certainly responsible for the relatively low calculated activation energy of the Na atoms diffusion. This mechanism demonstrates the interplay between atomic frameworks and ion-ion interactions in sustaining efficient long-range ion diffusion.

\section{Conclusions}
Structures of \ch{(TeO2)_{1-x}-(Na_{2}O)_{x}} glasses with x = 0.10, 0.1875, 0.30 and 0.40 were obtained by resorting to molecular modeling within 
the FPMD framework. The theoretically obtained X-ray total structure factors and total X-rays PDFs showed a fair agreement with the experimental
outcome. A reduction in Te coordination number occurs upon increase of the \ch{Na2O} content, corresponding to structural depolymerization of the glass network. This observed depolymerization originates from the replacement of Te-BO-Te by Te-NBO bonds. As a consequence, the concentration of NBO within this glassy binary system exhibits an upward trend strictly related to the concentration of the modifier, resulting in a decrease in the overall coordination number of Na. The provided analysis of the structural units distribution in terms of the parameter $Q_{m}^{n}$, 
for each chemical species, allows to rationalize each composition considered in this work and provides a quantitative support to 
both to the structural evolution of the different simulated systems and the contribution of Na$_2$O moieties to the depolymerization of the 
TeO$_2$ network.
Tentatively, we correlated the occurrence of the FSDP to the 
formation of channel-like structures in which Na$^+$ cations can diffuse. These channels are formed by Te and O atoms and are preserved at high temperatures by the rigidity of the Te sub-network. We demonstrated that the formation of these channels is mainly impacted by the sodium oxide content, rather than temperature. In terms of correlated dynamical effects, the ionic conductivity of the studied glasses increases upon increasing the concentration of sodium oxide.

\vspace{15mm}

\noindent\textbf{\Large Data availability}\\
Representative trajectory files of the four \ch{(TeO2)_{1-x}-(Na_{2}O)_{x}} with x = 0.10, 0.1875, 0.30 and 0.40 glasses will be available on GitHub: \textit{insert-link-here-before-publication}. Part of the structural analysis in this work was done using Amorphy suite of programs: \hyperlink{https://github.com/ASM2C-group/amorphy}{https://github.com/ASM2C-group/amorphy} \\

\noindent\textbf{\Large Acknowledgements}\\
This work was supported by the French ANR via the AMSES project (ANR-20-CE08-0021) and by r\'egion Nouvelle Aquitaine via the CANaMIAS project AAPR2021-2020-11779110. Calculations were performed by using resources from Grand Equipement National de Calcul Intensif (GENCI, projects No. 0910832, 0913426 and 0914978). We used computational resources provided by the computing facilities M\'esocentre de Calcul Intensif Aquitain (MCIA) of the Universit\'e de Bordeaux and of the Universit\'e de Pau et des Pays de l'Adour.

\bibliographystyle{apsrev4-2}
\bibliography{achemso-demo}

\pagebreak
\clearpage

\setcounter{equation}{0}
\setcounter{figure}{0}
\setcounter{table}{0}
\setcounter{page}{1}
\makeatletter
\renewcommand{\theequation}{S\arabic{equation}}
\renewcommand{\thefigure}{S\arabic{figure}}

\begin{table*}[!h]
\begin{center}
\textbf{\large Supplemental Materials: Atomic scale structure and dynamical properties of \ch{(TeO2)_{1-x}-(Na2O)_{x}} glasses through first-principles modeling  and XRD measurements}\\

Firas Shuaib, Assil Bouzid, Remi Piotrowski, Gaelle Delaizir, Pierre-Marie Geffroy, David Hamani, Raghvender Raghvender, Steve Dave Wansi Wendji, Carlo Massobrio, Mauro Boero, Guido Ori, Philippe Thomas, and Olivier Masson.
\end{center}
\end{table*}

\begin{figure*}[!htbp]
	\centering
\includegraphics[width=0.3\linewidth,keepaspectratio=true]{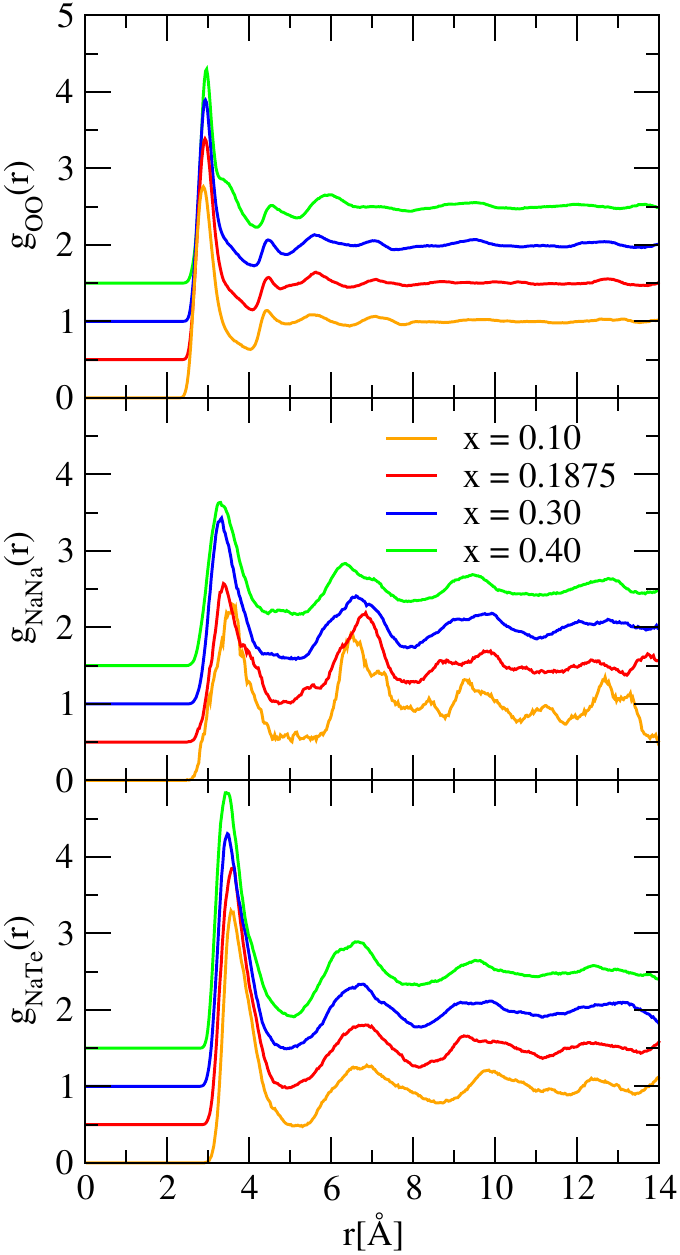}
	\caption{Te-Na, Na-Na and O-O partial pair distribution functions g$_{\alpha\beta}$ for  (TeO$_2$)$_{1-x}$-(Na$_{2}$O)$_{x}$ systems obtained from FPMD. The curves are shifted vertically for clarity.}  
	\label{fig_S0}
\end{figure*}

\begin{figure*}[!htbp]
	\centering
	\includegraphics[width=0.6\linewidth,keepaspectratio=true]{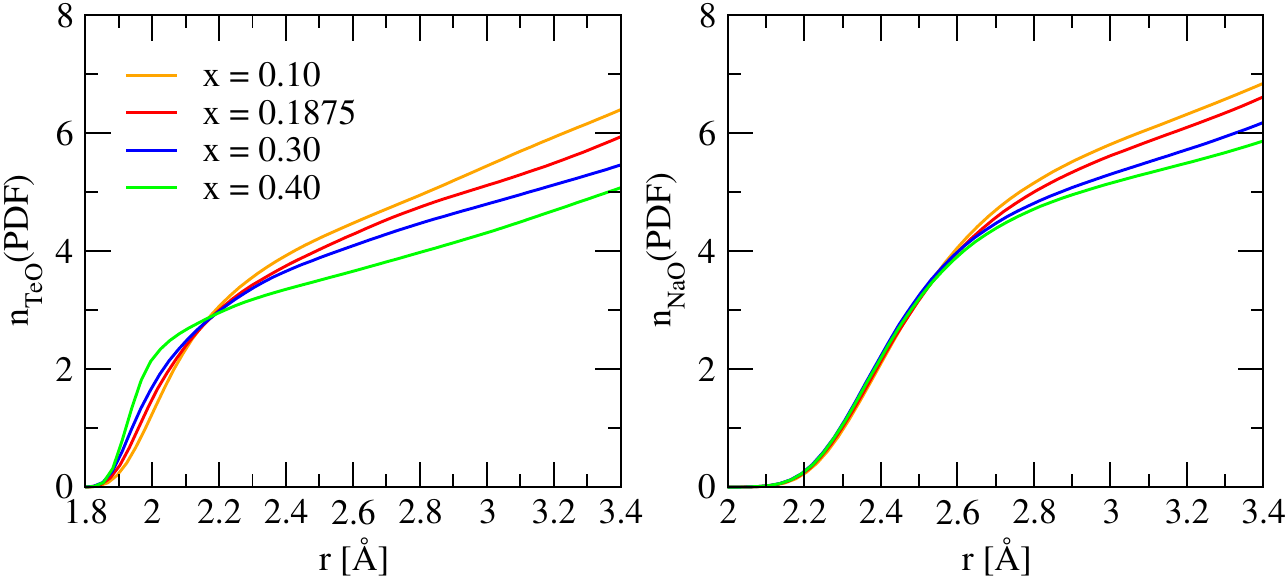}	
 \caption{The \ch{n_{Te-O}} (left panel) and \ch{n_{Na-O}} (right panel) coordination numbers as computed  by  integrating the partial pair distribution function within a cutoff distance  corresponding to its first minimum position for (TeO$_2$)$_{1-x}$-(Na$_{2}$O)$_{x}$ glasses at T = 300 K.}  
	\label{fig_S1}
\end{figure*} 

  \begin{figure*}[!htbp]
		\centering		\includegraphics[width=0.4\linewidth,keepaspectratio=true]{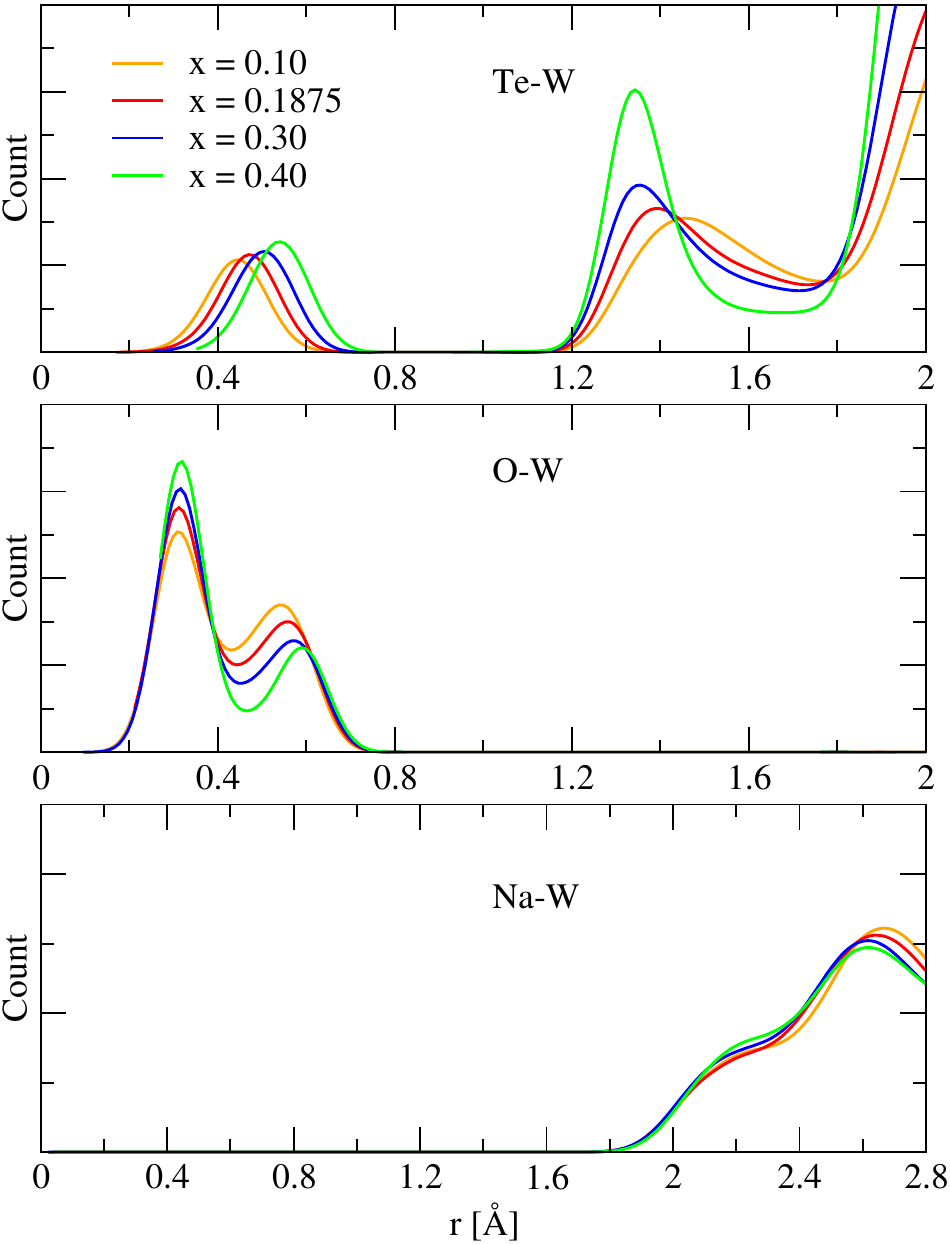}
  \caption{Distribution of the Wannier centers around \ch{Te}, \ch{O}, and \ch{Na} in(TeO$_2$)$_{1-x}$-(Na$_{2}$O)$_{x}$ glasses at T = 300 K.}  
	\end{figure*}

\begin{figure*}[!htbp]
		\centering
		\includegraphics[width=0.6\linewidth,keepaspectratio=true]{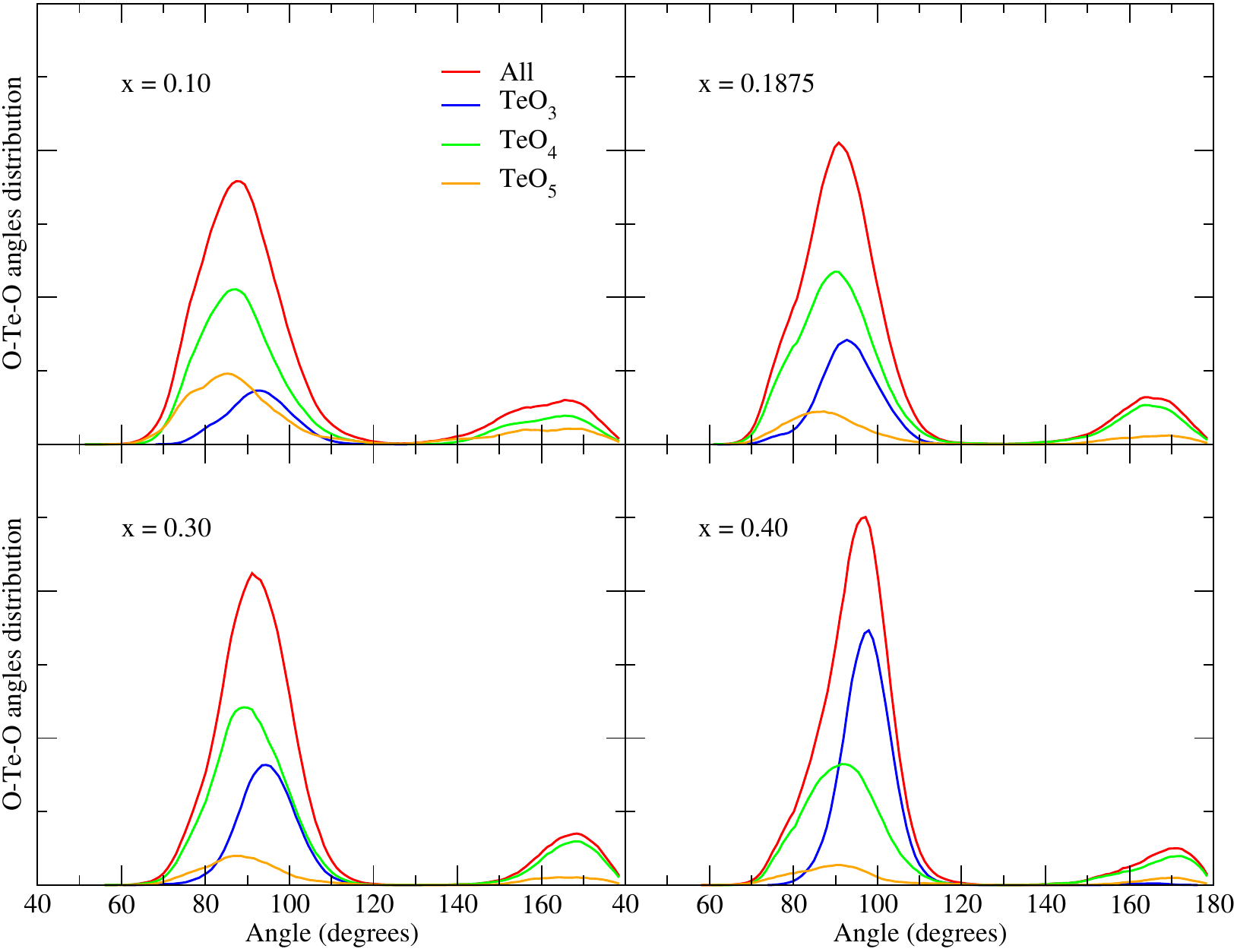}
  \caption{(Color online) Bond-angles distribution of O-Te-O angles in  (TeO$_2$)$_{1-x}$-(Na$_{2}$O)$_{x}$ glasses with x = 0.10, 0.1875, 0.30, 0.40. The O-Te-O angle distribution is splited into different parts. Those contributed by Te 3, 4, and 5 coordinated unit and then the bond-angles distribution of these parts are plotted.}  
	\end{figure*} 
 
 \begin{figure*}[!htbp]
		\centering
		\includegraphics[width=0.6\linewidth,keepaspectratio=true]{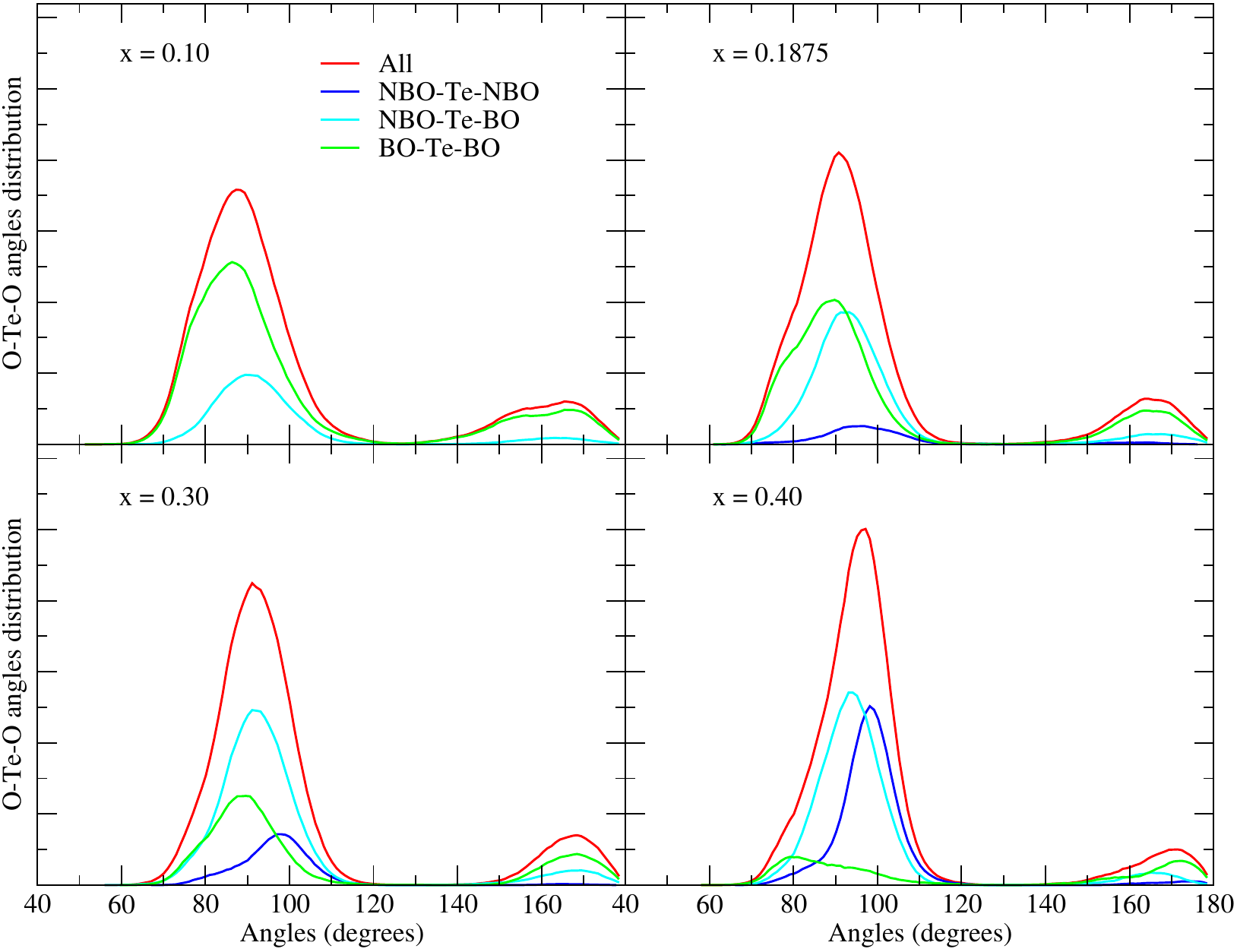}
  \caption{(Color online) O-Te-O bond-angle distribution with NBO-Te-NBO, NBO-Te-BO, BO-Te-BO angles contributions in  (TeO$_2$)$_{1-x}$-(Na$_{2}$O)$_{x}$ glasses with x = 0.10, 0.1875, 0.30, 0.40.}  
	\end{figure*} 

\begin{figure*}[!htbp]
	\centering
	\includegraphics[width=0.6\linewidth,keepaspectratio=true]{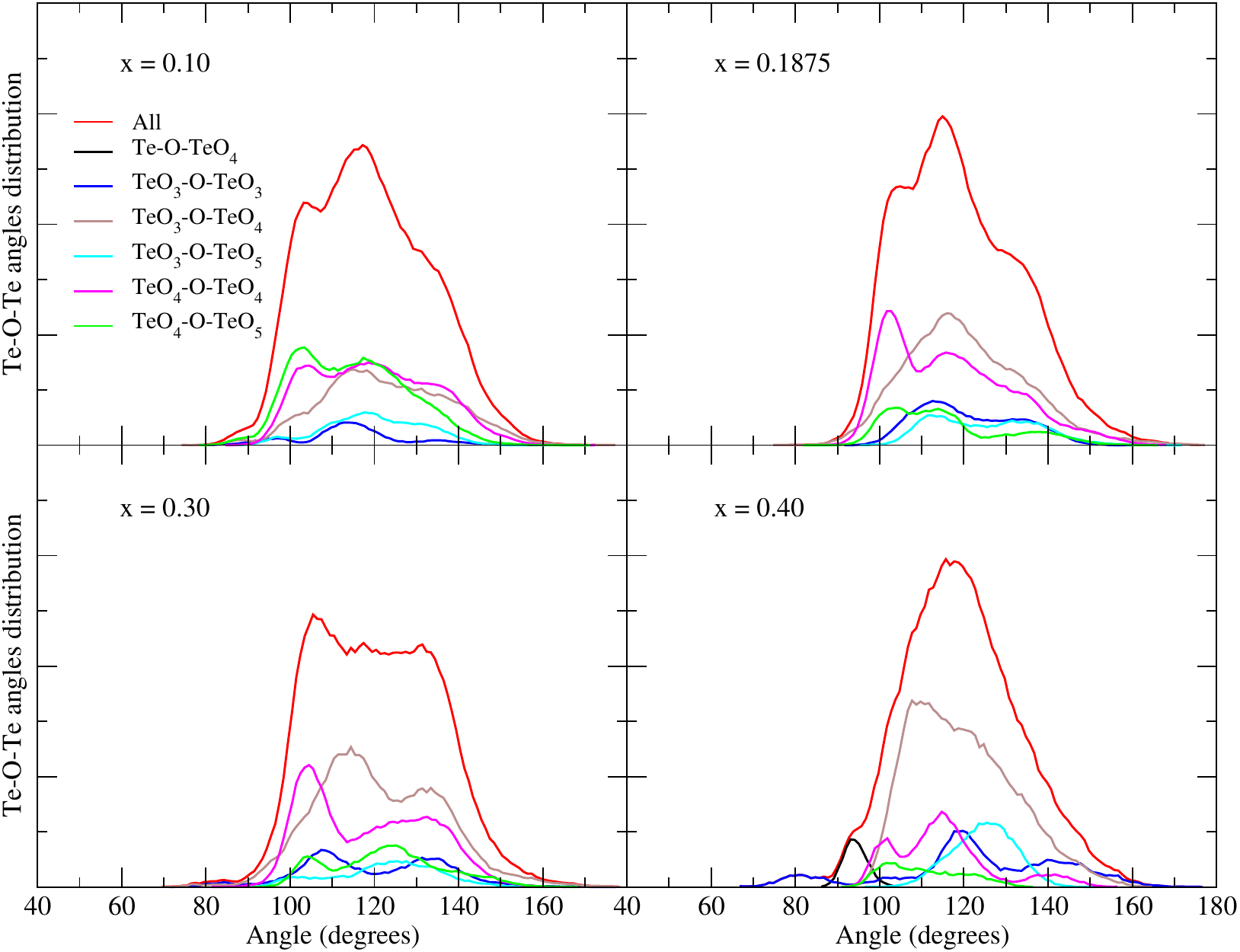}
	\caption{(Color online) Te-O-Te bond-angle distribution with considering contributions from all possible different local environments in Te chemical groups in (TeO$_2$)$_{1-x}$-(Na$_{2}$O)$_{x}$ glasses with x = 0.10, 0.1875, 0.30, 0.40.}  
\end{figure*}

\begin{figure*}[!htbp]
	\centering
	\includegraphics[width=0.6\linewidth,keepaspectratio=true]{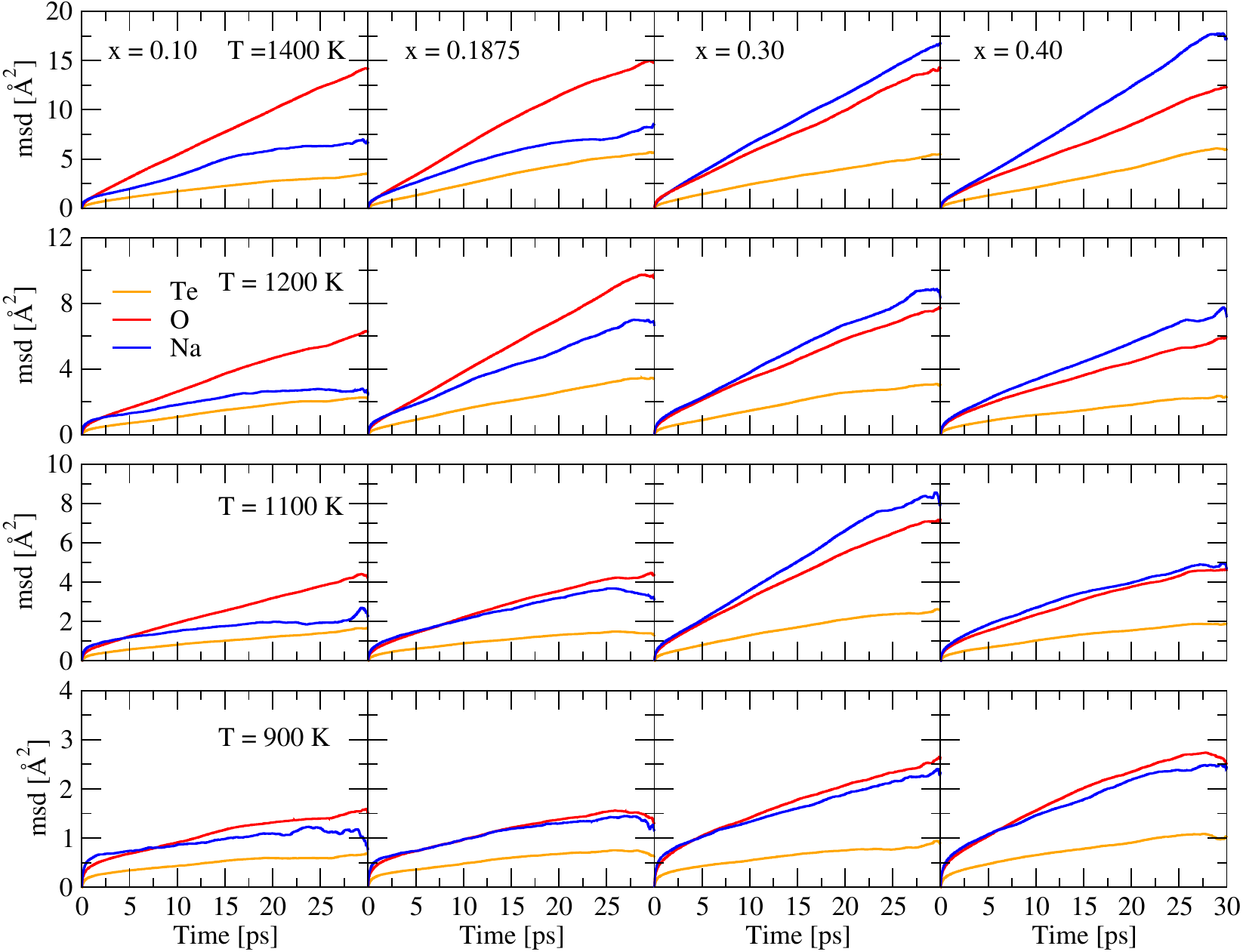}
	\caption{(Color online) Mean square displacements (MSD) of each element in the four (TeO$_2$)$_{1-x}$-(Na$_{2}$O)$_{x}$ systems at T = 1400 K, T = 1200 K, T = 1100 K, and T = 900 K. }  
\end{figure*}

\begin{figure*}[!htbp]
	\centering
	\includegraphics[width=0.6\linewidth,keepaspectratio=true]{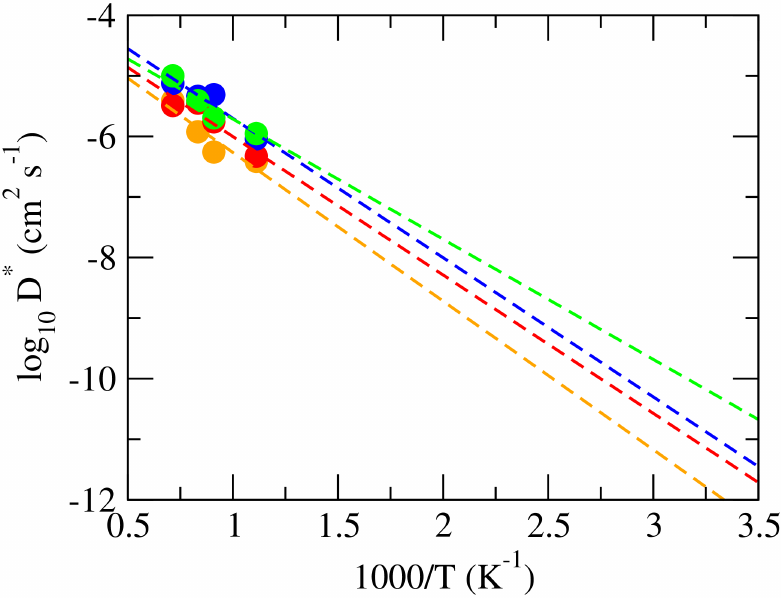}
	\caption{(Color online) Na-ion self-diﬀusion coeﬃcients (circles). Dashed lines correspond to a linear Arrhenius ﬁt that used to obtain the activation energies of Na ionic conduction for each (TeO$_2$)$_{1-x}$-(Na$_{2}$O)$_{x}$ glasses.}  
\end{figure*}

\begin{figure*}[!htbp]
	\centering
	\includegraphics[width=0.4\linewidth,keepaspectratio=true]{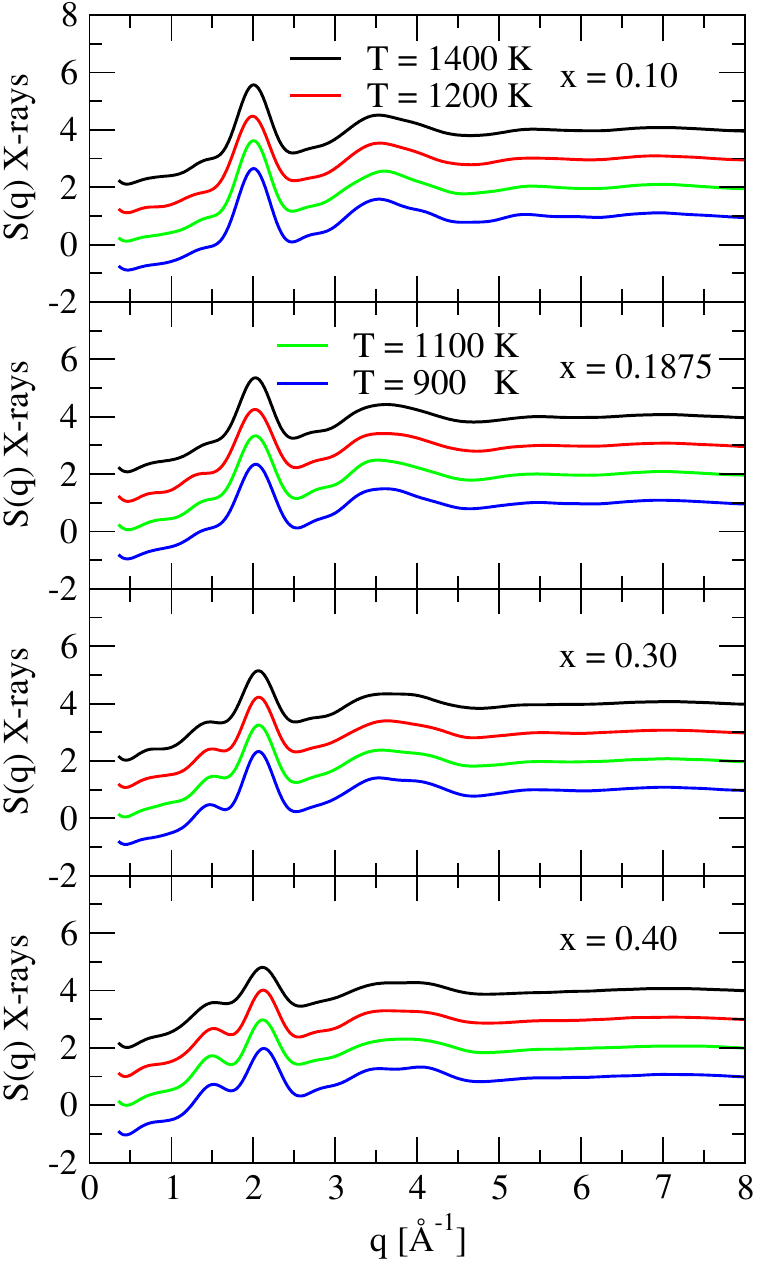}
	\caption{(Color online) Calculated total X-ray structure factor S$_{T}^{X}$(q) in the reciprocal space at the different target temperatures in  (TeO$_2$)$_{1-x}$-(Na$_{2}$O)$_{x}$ glasses with x = 0.10, 0.1875, 0.30, 0.40. Vertical shifts are applied for clarity.}  
\end{figure*}

\begin{figure*}[!htbp]
	\centering
	\includegraphics[width=0.99\linewidth,keepaspectratio=true]{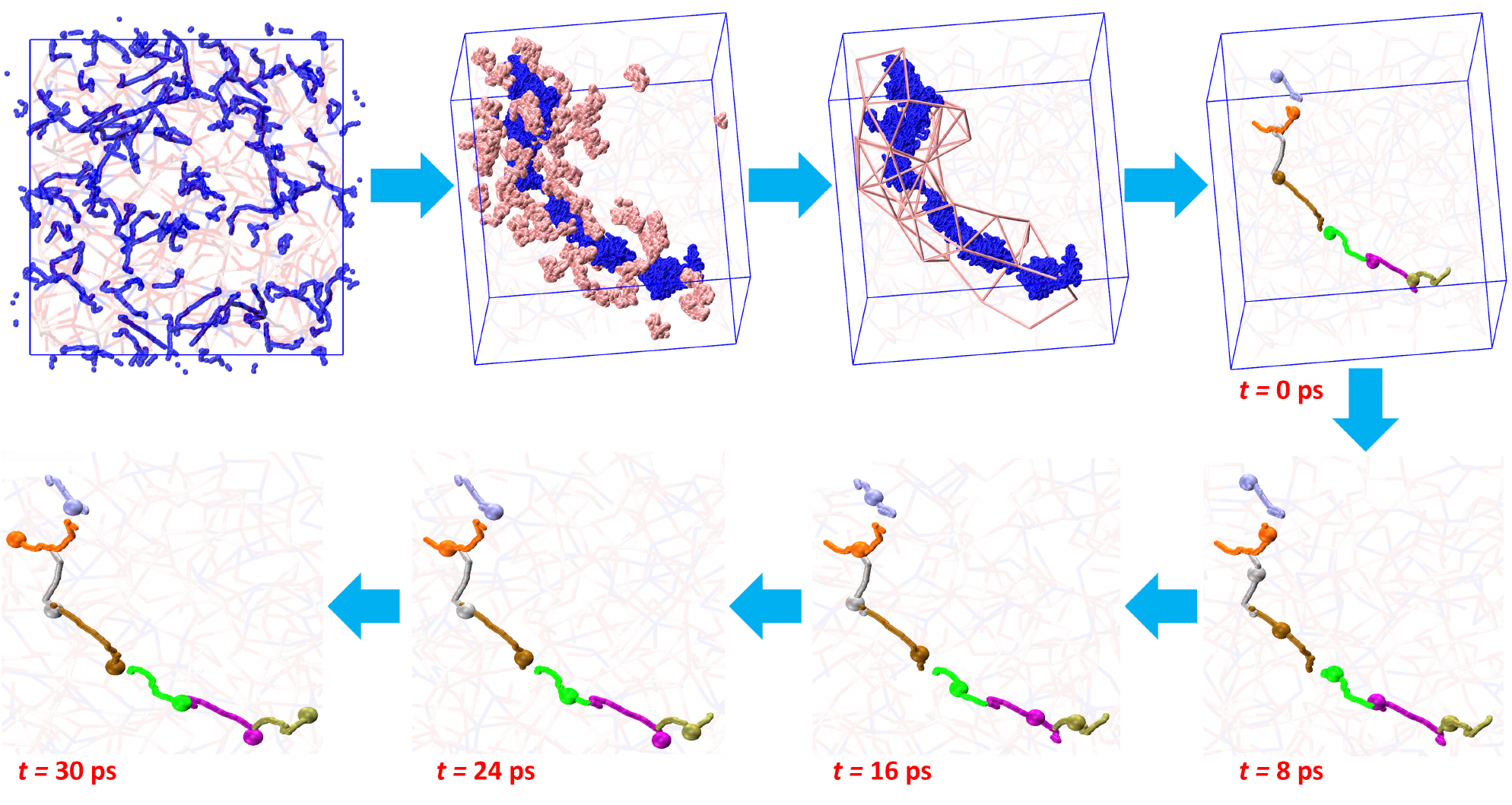}
	\caption{Accumulated trajectories over 30 ps at 1200 K. The blue spheres that are connected to each other represent the Na channels inside Te-Te bonds (Pink bonds), pink sphere represent the Te atoms for (TeO$_2$)$_{1-x}$-(Na$_{2}$O)$_{x}$ glasses with x = 0.30. O atoms are omitted for clarity. Other spheres with different colors that are connected to each other represent seven Na atoms that build one of the longest channels and their movement during the simulation time in the presented system.}  
\end{figure*}

\clearpage
\newpage



\end{document}